\documentclass{aa}

\usepackage[stable]{footmisc}
\usepackage{amsmath}
\usepackage[varg]{txfonts}
\usepackage{natbib}
\bibpunct{(}{)}{;}{a}{}{,}
\usepackage{graphicx}
\usepackage[breaklinks, colorlinks, citecolor=blue]{hyperref}
\usepackage{enumitem}
\usepackage[table,dvipsnames]{xcolor}
\usepackage{fontawesome5}

\def\lsim{\mathrel{\raise .4ex\hbox{\rlap{$<$}\lower 1.2ex\hbox{$\sim$}}}}
\def\gsim{\mathrel{\raise .4ex\hbox{\rlap{$>$}\lower 1.2ex\hbox{$\sim$}}}}
\renewcommand{\d}{\text{d}}
\def\muk{$(\mu{\rm K})^2$ }
\newcommand{\ghz}{\,{\rm GHz}}
\newcommand{\Gyr}{{\rm Gyr}}
\newcommand{\lcdm}{$\Lambda$CDM}
\newcommand{\planck}{\textit{Planck}}
\newcommand{\camb}{{\tt camb}}
\newcommand{\um}{\texttt{UniverseMachine}}
\newcommand{\coco}{CO$\times$CO}
\newcommand{\cocib}{CO$\times$CIB}
\newcommand{\lact}{ACT-like}
\newcommand{\lcmnhd}{CMB-HD-like}

\setlist[itemize]{label=\textbullet}

\begin{document}

\title{Extragalactic carbon-monoxide (CO) as a CMB foreground: an analytic halo model for the CO auto-spectrum and its correlation with the Cosmic Infrared Background}

\author{
G. Zagatti\inst{1,2,3}~\thanks{\email{giorgia.zagatti@unife.it}}
\and
M. Douspis\inst{3}
\and
G. Fabbian\inst{3}
\and
L. Salvati\inst{3}
}

\institute{\small Dipartimento di Fisica e Scienze della Terra, Universit\`a degli Studi di Ferrara, via Saragat 1, I-44122 Ferrara, Italy\goodbreak
\and
Istituto Nazionale di Fisica Nucleare, Sezione di Ferrara, via Saragat 1, I-44122 Ferrara, Italy\goodbreak
\and
Universit\'{e} Paris-Saclay, CNRS, Institut d'Astrophysique Spatiale, 91405 Orsay, France}

\date{Received XXX; accepted YYY}

\abstract{\textit{Context}. Carbon Monoxide (CO) lines emitted by star-forming galaxies are redshifted into the frequency bands of the Cosmic Microwave Background (CMB) experiments. Their correlation with the Cosmic Infrared Background (CIB) may be a non-negligible small-scale foreground.\\
\textit{Aims.} We develop an analytic, full-shape halo model framework for the CO auto-spectrum and for the \cocib\,cross-spectrum in CMB bands, and assess its impact on current and future high-resolution CMB experiments.\\
\textit{Methods}. We use a luminosity-weighted halo model with separate occupation distributions of central and satellite galaxies for CO and CIB emitters, CO transitions up to $J=13$, and all auto- and cross-frequency spectra. Line and infrared luminosities derive from the same star formation rate. We perform Fisher analysis for \lact\,and \lcmnhd\,experiments.\\
\textit{Results.} At $\ell=3000$, the \cocib\,contribution to the sky power is 0.4, 0.7, and 1.9\muk\,at 90, 150, and 220\ghz, and the CO auto-spectrum is $\simeq0.02$\muk. For an \lact\,experiment, the CO occupation parameters are essentially unconstrained, and neglecting CO leaves the cosmological parameters unbiased, but shifts the radio and SZ parameters by $\sim2\sigma$ and $\simeq1\sigma$, respectively. For a \lcmnhd\,experiment, CO is detected with high significance and the minimum halo masses of CO and CIB emitters are constrained; neglecting CO biases the cosmological parameters up to $\sim2\sigma$ when all multipoles are used in the analysis.\\
\textit{Conclusions.} CO should be included in the foreground model of future high-resolution CMB analyses; our framework describes it analytically and consistently with the CIB, and the code implementing it is publicly available.
}


\authorrunning{Zagatti et al.}

\titlerunning{Extragalactic CO as a CMB foreground: an analytic HM for the CO auto-spectrum and its correlation with the CIB}

\maketitle
\section{Introduction}\label{sec:intro}
    On angular scales smaller than a few arcminutes, the temperature power spectrum measured by high-resolution Cosmic Microwave Background (CMB) experiments is no longer dominated by the primary CMB anisotropies but by extragalactic foregrounds and by secondary anisotropies other than gravitational lensing. These are the Cosmic Infrared Background (CIB) from dusty star-forming galaxies, the thermal and kinetic Sunyaev--Zel'dovich (tSZ and kSZ) effects, their correlations (in particular tSZ$\times$CIB), and the emission of radio point sources~\citep[e.g.][]{Dunkley:2013vu, George:2014oba, Reichardt:2021, Louis:2025}. The accuracy of the cosmological parameters inferred from small scales therefore depends on how well these components are modelled.
    This becomes critical for the current and upcoming generations of experiments, such as the Simons Observatory~\citep{SimonsObservatory:2018koc} and CMB-HD~\citep{Sehgal:2019,CMBHD:2022}, which will measure the temperature power spectrum up to $\ell\sim10^4$ and beyond. Recent joint analyses of \planck, ACT and SPT data with a common foreground model show that the \lcdm\,parameters are stable against changes in the foreground modelling, while opening up the parameter space to extensions such as the sum of the neutrino masses, $\sum m_\nu$, and the effective number of relativistic species, $N_{\rm eff}$, makes the constraints more sensitive to the choice of foreground model~\citep{Tristram:2025}.

    Some extragalactic emission is not included in any of these foreground models. Star-forming galaxies emit not only thermal dust continuum but also a number of bright far-infrared and sub-millimetre emission lines. Among them, the rotational transitions of carbon monoxide (CO) are the most widely used tracers of cold molecular gas, which is the fuel of star formation~\citep[see e.g.][for reviews]{Carilli:2013, Bolatto:2013}. The rotational ladder of CO is spaced by $\simeq115$\ghz\,in the rest frame, and lines emitted at $0\lsim z\lsim 6$ are redshifted into the frequency bands commonly used by CMB experiments. CO line emission is the target of Line Intensity Mapping (LIM) experiments, which use narrow spectral channels to reconstruct the three-dimensional distribution of the unresolved line emission \citep[e.g.][]{Visbal2010,Pullen:2013,Breysse:2014,Keating:2016}; see~\cite{Kovetz:2017} and~\cite{Bernal:2022} for reviews.

    \cite{Righi:2008} first pointed out that extragalactic CO acts as a foreground for CMB observations. More recently,~\cite{Maniyar:2023} and~\cite{Kokron:2024}, using simulated galaxy catalogues and empirical relations between infrared and CO luminosities, showed that the auto-spectrum of the CO emission is small, but its correlation with the CIB may not be negligible, and may even be comparable with extragalactic signals currently included in CMB analyses. In particular, they found that at $\ell\simeq3000$ and at 150 and 220\ghz, its amplitude is comparable to the kSZ and the tSZ$\times$CIB power, and that neglecting CO can therefore bias the kSZ amplitude inferred from the CMB temperature power spectrum. \cite{Mehta:2025} found that neglecting \cocib\,in a CMB-S4-like analysis would bias foreground parameters by many standard deviations. On the observational side,~\cite{Roy:2024CO32} reported the first measurement of extragalactic CO in CMB data, by stacking \planck\,maps on the positions of eBOSS galaxies to measure the mean $\rm{CO}(3\rightarrow2)$ intensity at $z\simeq0.5$.

    Understanding the CO emission is therefore crucial in the analysis of future high-resolution CMB data. However, the available predictions rely mainly on simulations, while a coherent analytic framework is currently missing. In this study, we develop such a framework. Our work is in the spirit of modelling the small-scale foregrounds and secondary anisotropies with physically motivated models, rather than marginalising over templates with free amplitudes, as done for the tSZ effect~\citep{Douspis:2021ing}, the KSZ effect~\citep{Gorce:2022cvb}, and the CIB~\citep{Zagatti:2024bhs}. Our goal is to provide a flexible analytic model of the CO auto-spectrum and of the \cocib\,cross-spectrum in CMB bands, physically connected to the star formation rate (SFR) of galaxies and to their host dark matter haloes, which can be evaluated for any set of frequency bands and multipoles. The code implementing the model is publicly available\footnote{\href{https://github.com/giorgiazagatti/extragalacticCO_CIB.git}{\color{magenta}\faGithub\,\, giorgiazagatti/extragalacticCO\_CIB}}.

    The paper is organised as follows. In Sect.~\ref{sec:eCO}, we review the physics of extragalactic CO emission and the way it enters CMB observations. In Sect.~\ref{sec:model}, we present the halo model for the \coco\,and the \cocib\,power spectra. In Sect.~\ref{sec:mod_ass}, we test the impact of the main model assumptions at the power-spectrum level. In Sect.~\ref{sec:results}, we present the results of the Fisher analysis for \lact\,and \lcmnhd\,setups and quantify the bias induced by neglecting CO. In Sect.~\ref{sec:discussion}, we compare the predictions of the model with the other components of the small-scale CMB sky and with previous works, and discuss the implications of our results for CMB analyses and the limitations of the model. We conclude in Sect.~\ref{sec:conc}.

\section{Extragalactic CO}\label{sec:eCO}
    CO is a diatomic molecule that radiates through rotational transitions at specific emission frequencies $\nu_J \simeq J\nu_1$. Here $J$ indicates the rotational quantum number and $\nu_1=115.3$\ghz\,is the rest-frame frequency of the ground transition, $\rm{CO}(1\rightarrow0)$.

    In galaxies, CO resides in cold molecular clouds together with molecular hydrogen, H$_2$. Although H$_2$ dominates the mass of molecular gas, it has no permanent dipole moment and its lowest rotational transition lies $E/k_B\simeq510$\,K above the ground state, so it does not radiate at the typical temperatures of molecular clouds ($\sim10$--$50$\,K). In contrast, CO is the second most abundant molecule after $H_2$, and its low-$J$ levels ($E/k_B\simeq5.5$\,K for $J=1$) are easily excited. For this reason, it is the most widely used tracer of molecular gas and, since molecular gas fuels star formation, of the star formation activity of galaxies \citep{Carilli:2013, Bolatto:2013}.

    Extragalactic CO can be observed in two main regimes: the resolved emission from individual galaxies and the unresolved emission, in which individual galaxies are not detected and one measures the fluctuations of the cumulative line intensity. This work focuses on the latter regime. At high redshift in particular, galaxies are typically unresolved, and the gas properties can only be constrained on average through the global CO emission of the galaxy population. The unresolved CO emission traces the large-scale structure (LSS) and the connection between galaxies and their host dark matter haloes; it is both a cosmological signal and a foreground for CMB experiments.

\begin{figure}
        \centering
        \includegraphics[width=0.85\linewidth]{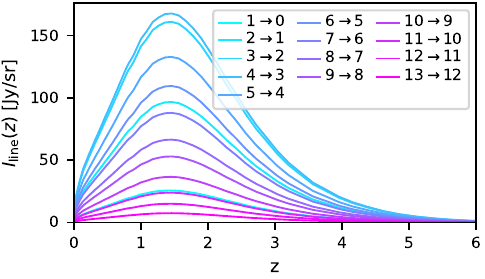}
        \caption{Mean line intensity as a function of emission redshift for all the CO transitions up to $J_{\rm max}=13$ in the baseline model (\um\ SFR and \cite{Visbal2010} SLED, normalised as in Sect.~\ref{subsec:co_lum}).}
        \label{fig:line_int}
    \end{figure}

    The mean specific intensity of a line with rest-frame frequency $\nu_{\rm rest}$ emitted at redshift $z$ is
    \begin{equation}\label{eq:line_int}
        I_{\rm line}(z)=\dfrac{c}{4\pi}\dfrac{1}{\nu_{\rm rest}H(z)}\int {\rm d}M\,\dfrac{{\rm d}n}{{\rm d}M}(M,z)\,L_{\rm line}(M,z) ,
    \end{equation}
    where $H(z)$ is the Hubble parameter, $L_{\rm line}(M,z)$ is the total luminosity of a halo of mass $M$ (see Sect.~\ref{subsec:co_lum}), and $\d n/\d M$ is the halo mass function (details in Sect.~\ref{subsec:hm}). Figure \ref{fig:line_int} shows the line intensity predicted by our baseline model, i.e. the reference set of model choices used throughout this work (\um\,SFR and the CO spectral line energy distribution, SLED, of~\cite{Visbal2010}), for the CO ladder from $\rm{CO}(1\rightarrow0)$ to $\rm{CO}(13\rightarrow12)$. Since the baseline CO SLED (Sect.~\ref{subsec:co_sled}) does not depend on redshift, all transitions share the same redshift dependence and peak at $z\simeq1.3$, at redshift lower than the peak of the cosmic star formation rate density~\citep[$z\simeq2$,][]{Madau:2014}, because of the factor $1/H(z)$ of Eq.~\eqref{eq:line_int}, which decreases with redshift. The curves of the different transitions therefore have the same shape and differ only in amplitude. From Eq.~\eqref{eq:line_int}, $I_{\rm line}\propto L_{\rm line}/\nu_{\rm rest}$, with $L_{\rm line}=R_J\rm{SFR}$ and $R_J$ the coefficient of the CO SLED for the transition J (Sect.~\ref{subsec:co_sled}), so that the amplitude scales as $R_J/\nu_J$. This ratio is largest for the mid-$J$ transitions, $J=3-5$, which are the brightest, and decreases with $J$ above $J=4$.

    Redshift moves each line to the observed frequency $\nu_{\rm obs}=\nu_{\rm rest}/(1+z)$. CMB experiments have wide bandpasses, with a typical fractional width of $\Delta\nu/\nu\sim0.3$; therefore, a band spanning $[\nu_{\rm min}, \nu_{\rm max}]$ collects photons from CO transitions over the redshift range
    \begin{equation}\label{eq:zrange}
        \dfrac{\nu_J}{\nu_{\rm max}}-1\ \leq\ z\ \leq\ \dfrac{\nu_J}{\nu_{\rm min}}-1 .
    \end{equation}
    The rest-frame frequencies $\nu_J$ of all transitions considered in this work are listed in Table 1 of~\cite{Visbal2010}. For example, the ACT 150\ghz\ band \citep[124--172\ghz,][]{Louis:2025} receives CO$(2\rightarrow1)$ from $0.34\leq z\leq0.86$, CO$(3\rightarrow2)$ from $1.01\leq z\leq1.79$, and CO$(4\rightarrow3)$ from $1.68\leq z\leq2.72$. Each band therefore collects several transitions from different, partially overlapping redshift intervals, and CO in CMB bands behaves more like a continuum source than like a single narrow line. For this reason, it cannot be removed as an isolated spectral feature and has to be modelled consistently within multi-frequency foreground analyses.

\section{The model}\label{sec:model}
   In this work, we build an analytic, full-shape halo model framework for the extragalactic CO power spectrum in CMB bands, with an explicit dependence on the parameters of a halo occupation distribution (HOD). This approach connects the CO emission directly to galaxy populations and makes it possible to predict, within the same framework, the cross-correlations with the CIB and with other extragalactic components\footnote{In this work, we only consider the \cocib\,cross-correlation since CO$\times$tSZ is expected to be subdominant~\citep{Maniyar:2023, Kokron:2024}. We leave CO$\times$tSZ for future works.}. Throughout this work, galaxies populating dark matter haloes are divided into central and satellite galaxies \citep[e.g.][]{Zheng:2005, Tinker:2009mx}, with the mean numbers of those galaxies in a halo of mass \textit{M} given by
   \begin{equation}\label{eq:N_cent}
        \langle N_{\text{cent}}|M \rangle = \dfrac{1}{2}\Biggl[1+\text{erf}\Biggl(\dfrac{\log M-\log M_{\text{min}}}{\sigma_{\log M}}\Biggr)\Biggr] ,
    \end{equation}
    and
    \begin{equation}\label{eq:N_sat}
        \langle N_{\text{sat}}|M \rangle = \dfrac{1}{2}\Biggl[1+\text{erf}\Biggl(\dfrac{\log M-\log2M_{\text{min}}}{\sigma_{\log M}}\Biggr)\Biggr]\Biggl(\dfrac{M}{M_{\text{sat}}}\Biggr)^\alpha .
    \end{equation}
    Here, halo masses are in $M_\odot\,h^{-1}$, $M_{\rm min}$ is the characteristic minimum mass of haloes hosting a central galaxy, $\sigma_{\log M}$ is the width of the transition, $M_{\rm sat}$ is the mass scale at which a halo hosts on average one satellite (for $M_{\rm sat}\gg M_{\rm min}$), and $\alpha$ is the slope of the satellite occupation. Satellites are switched on at $2M_{\rm min}$, so that haloes host a central before hosting satellites. We use independent sets of HOD parameters for CO emitters and for CIB emitters. Their fiducial values are given in Table~\ref{tab:fg_params}\footnote{As detailed in Appendix~\ref{app:um_sfr}, the satellite occupation counts effective, SFR-weighted emitters rather than galaxies.}.

    In the following, after a brief review of the halo model, we describe its implementation for the extragalactic \coco\,power spectrum and then extend it to the \cocib\,cross-correlation.

    \subsection{Halo model basics}\label{subsec:hm}
        The main building block of our framework is the halo model. Here, we recall its basic concepts and refer the reader to \cite{Cooray:2002dia} and \cite{Asgari:2023mej} for comprehensive reviews.

        The halo model assumes that all matter, and any tracer field, can be described as the sum of contributions from virialised dark matter haloes. In Fourier space, the auto- or cross-power spectrum of two fields \textit{u} and \textit{v} is decomposed as
        \begin{equation}
          P_{uv}(k)=P^{1h}_{uv}(k)+P^{2h}_{uv}(k),
        \end{equation}
        where the one-halo (1h) term,
        \begin{equation}\label{eq:1hterm}
          P^{1h}_{uv}(k)=\int_0^\infty{\rm d}M\,\frac{{\rm d}n}{{\rm d}M}(M)\,\hat W_u(M,k)\,\hat W_v(M,k),
        \end{equation}
        describes correlations within the same halo, and the two-halo (2h) term,
        \begin{equation}\label{eq:2hterm}
          P^{2h}_{uv}(k)
          =P^{\rm lin}_{mm}(k)\,
           \prod_{x=u,v}\left[\int_0^\infty{\rm d}M\,\frac{{\rm d}n}{{\rm d}M}(M)\,b(M)\,\hat W_x(M,k)\right],
        \end{equation}
        describes correlations between distinct haloes. Here, $P_{\rm mm}^{\rm lin}$ is the linear matter power spectrum, and $\hat{W}(M,k)=W(M)\tilde{u}(M,k)$ separates the amplitude of the tracer, $W(M)$, from its normalised profile in Fourier space, $\tilde{u}(M,k)$.

        The key ingredients of the halo model are the halo mass function, the halo bias, and the halo density profile. The halo mass function $\d n/\d M$ is the comoving number density of haloes of mass \textit{M} at redshift $z$, for which we use the fitting formula of \cite{Tinker:2008ff}. The halo bias $b(M)$ relates the halo overdensity to the matter overdensity on large scales, and we adopt the prescription of \cite{Tinker:2010my}. Finally, we adopt a Navarro--Frenk--White (NFW) profile \citep{Navarro:1996gj} for the radial distribution of matter, and of satellite galaxies, within haloes, with the concentration--mass relation $c(M,z)=9\,(1+z)^{-1}(M/M_\star)^{-0.13}$~\citep{Bullock:1999he}, where $M_\star(z)$ is the non-linear mass, defined by $\sigma(M_\star,z)=\delta_c=1.686$, with $\sigma(M,z)$ the rms of the linear density field. Haloes are defined with an overdensity of 200 with respect to the critical density.

        With these ingredients, the halo model predicts the clustering of matter and of its tracers through Eqs.~\eqref{eq:1hterm} and~\eqref{eq:2hterm}. The angular power spectrum of two fields \textit{u} and \textit{v}, observed in the frequency channels $\nu$ and $\nu^\prime$, is then obtained in the Limber approximation \citep{Limber:1953, LoVerde:2008} as
        \begin{equation}\label{eq:ps_generic}
            C_{\ell,\nu\times\nu^\prime}^{uv} = \int{\rm d}z\,\dfrac{{\rm d}\chi}{{\rm d}z}\dfrac{1}{\chi^2}\,\xi_\nu^u(z)\,\xi_{\nu^\prime}^v(z)\,P_{uv}\!\left(k=\dfrac{\ell+1/2}{\chi},z\right) ,
        \end{equation}
        where $\chi(z)$ is the comoving distance, and $\xi_\nu^{u,v}(z)$ are the emission kernels of the tracers under study.

        \subsection{\coco\,power spectrum}\label{subsec:coco}
            To compute the CO power spectrum in a full-shape halo model framework, we need a luminosity-weighted power spectrum. Unlike in halo model analyses of the CIB \citep[e.g.][]{Maniyar:2020tzw,Zagatti:2024bhs}, the signal in a channel cannot be described as a single tracer with one emission kernel: at a given observed frequency the CO field in a channel is the sum of distinct lines, and its power spectrum is a sum over pairs of transitions. In the following, we present the general structure of the model for the CO auto-spectrum. The different ingredients and assumptions entering the modelling are discussed in Sect.~\ref{sec:mod_ass}.

            The contribution of the transitions $J$ and $J^\prime$ to the CO power spectrum between the channels $\nu$ and $\nu^\prime$ is
            \begin{align}\label{eq:clCO_single}
                C_{\ell,\nu\times\nu^\prime}^{{\rm CO}(J)\times {\rm CO}(J^\prime)} &= \int{\rm d}z\,\dfrac{{\rm d}\chi}{{\rm d}z}\dfrac{1}{\chi^2}\,\xi_\nu^{J}(z)\,\xi_{\nu^\prime}^{J^\prime}(z)\\
                &\times P_{{\rm CO}(J)\times {\rm CO}(J^\prime)}\!\left(k=\dfrac{\ell+1/2}{\chi},z\right) ,
            \end{align}
            where
            \begin{itemize}
              \item $\xi_\nu^J(z)$ is the emission kernel of the transition $J$ in the channel $\nu$,
              \begin{equation}\label{eq:emiss_kernel}
                    \xi_\nu^J(z) = \dfrac{c}{4\pi}\dfrac{1}{H(z)\,\Delta\nu\,(1+z)^2}\,\overline{j}_J(z)\,\mathcal{R}_{\nu}\!\left(\dfrac{\nu_{J}}{1+z}\right) ,
                \end{equation}
                with $\Delta\nu$ the width of the channel, $\mathcal{R}_\nu$ its spectral response (a top-hat between the channel edges) evaluated at the observed frequency of the line, and $\overline{j}_J$ the mean comoving luminosity density defined in Eq.~\eqref{eq:jbar} and detailed in Appendix~\ref{app:kernels};
              \item $P_{\rm CO(J)\times\rm CO(J^\prime)}(k,z)$ is the three-dimensional CO power spectrum computed in the halo model.
            \end{itemize}
            To compute $P_{\rm CO(J)\times\rm CO(J^\prime)}(k,z)$, we define the two generic fields entering Eqs. \eqref{eq:1hterm} and \eqref{eq:2hterm} as
            \begin{equation}
                \begin{aligned}
                    \hat W_u(M,k)&\equiv \hat W_{{\rm CO}(J)}(M,k)=\dfrac{\langle L_c^J\rangle+\langle L_s^J\rangle\,\tilde{u}_{\rm gal}(k,M)}{\overline{j}_J},\\
                    \hat W_v(M,k)&\equiv \hat W_{{\rm CO}(J^\prime)}(M,k)=\dfrac{\langle L_c^{J^\prime}\rangle+\langle L_s^{J^\prime}\rangle\,\tilde{u}_{\rm gal}(k,M)}{\overline{j}_{J^\prime}} .
                \end{aligned}
            \end{equation}
            In these equations:
            \begin{itemize}
                \item $\langle L_{\rm c/s}^J\rangle$ is the mean CO luminosity of the central/satellite galaxies in a halo of mass \textit{M},
                \begin{equation}\label{eq:Lcs}
                    \langle L_{c/s}^J\rangle(M,z) = \langle N_{c/s}\rangle\,\ell^J_{c/s}(M,z),
                \end{equation}
                where $\ell^J_{c/s}(M,z)=R_{J}\,{\rm SFR}_{c/s}(M,z)$. Here, $\langle N_{\rm c/s}\rangle$ is the mean number of central/satellite galaxies given by Eqs.~\eqref{eq:N_cent} and~\eqref{eq:N_sat}, $\ell_{\rm c/s}^J$ is the CO luminosity of a single central/satellite galaxy, $R_J$ is the coefficient of the CO SLED that converts SFR into line luminosity (see Sect.~\ref{subsec:co_sled}), and ${\rm SFR}_{\rm c/s}$ is the star formation rate of a single central/satellite galaxy, which in our baseline model is derived from \um\,\citep{Behroozi:2019}, as described below;
                \item $\overline{j}_J$ is the mean emissivity of the transition $J$,
                \begin{equation}\label{eq:jbar}
                    \overline{j}_J(z) \equiv\int {\rm d}M\,\dfrac{{\rm d}n}{{\rm d}M}\bigl[\langle L_c^J\rangle+\langle L_s^J\rangle\bigr] ;
                \end{equation}
                \item $\tilde{u}_{\rm gal}(k,M)$ is the Fourier transform of the (normalised) spatial distribution of satellite galaxies within the halo, for which we assume an NFW profile. Central galaxies sit at the halo centre.
            \end{itemize}
            In our baseline model, $\rm{SFR}(M,z)$ is the mean total SFR of a halo of mass $M$, as given by \um\,(Sect.~\ref{subsec:co_sfrs}). We split it between central and satellite galaxies using the satellite fraction of \um, $f_{\rm sat}=\rm{SFR}_{s,\rm tot}^{\rm UM}/(\rm{SFR}_{c}^{\rm UM}+\rm{SFR}_{s,\rm tot}^{\rm UM})$, where $\rm{SFR}_{c}^{\rm UM}$ and $\rm{SFR}_{s,\rm tot}^{\rm UM}$ are the mean SFRs of the central galaxy and of all the satellites of the halo, respectively.

            Inserting these definitions into the halo model expressions, we obtain the 1h, the 2h, and the shot noise terms as
            \begin{equation}\label{eq:P1h_coco}
                \begin{aligned}
                    P_{{\rm CO}(J)\times {\rm CO}(J^\prime)}^{1h}&=\dfrac{1}{\overline{j}_J\,\overline{j}_{J^\prime}}\int {\rm d}M\,\dfrac{{\rm d}n}{{\rm d}M}\biggl[N_cN_s\Bigl(\ell_c^J\ell_s^{J^\prime}+ \ell_s^J\ell_c^{J^\prime}\Bigr)\tilde{u}_{\rm gal}\\
                    & + N_s^2\,\ell_s^J\ell_s^{J^\prime}\,\tilde{u}_{\rm gal}^2\biggr] ,
                \end{aligned}
            \end{equation}
            \begin{equation}\label{eq:P2h_coco}
                \begin{aligned}
                    P_{{\rm CO}(J)\times {\rm CO}(J^\prime)}^{2h} &= P_{mm}^{\rm lin}(k)\biggl[\dfrac{1}{\overline{j}_J}\int {\rm d}M\,\dfrac{{\rm d}n}{{\rm d}M}\,b(M)\bigl(N_c\ell_c^J+N_s\ell_s^J\tilde{u}_{\rm gal}\bigr)\biggr]\\
                    &\times\biggl[\dfrac{1}{\overline{j}_{J^\prime}}\int {\rm d}M\,\dfrac{{\rm d}n}{{\rm d}M}\,b(M)\bigl(N_c\ell_c^{J^\prime}+N_s\ell_s^{J^\prime}\tilde{u}_{\rm gal}\bigr)\biggr] ,
                \end{aligned}
            \end{equation}
            \begin{equation}\label{eq:Pshot_coco}
                P_{{\rm CO}(J)\times {\rm CO}(J^\prime)}^{\rm shot} = \dfrac{1}{\overline{j}_J\,\overline{j}_{J^\prime}}\int {\rm d}M\,\dfrac{{\rm d}n}{{\rm d}M}\biggl[N_c\,\ell_c^J\ell_c^{J^\prime}+N_s\,\ell_s^J\ell_s^{J^\prime}\biggr] .
            \end{equation}
            The shot noise term accounts for the emission of the same galaxy in the two transitions (galaxy pairs at zero separation). To simplify the notation, we have written $N_{\rm c/s}\equiv \langle N_{\rm c/s}\rangle$ and dropped the $(M,z)$ dependence. In the 1h term, we assume that the number of satellites in a halo follows a Poisson distribution, so that $\langle N_s(N_s-1)\rangle=\langle N_s\rangle^2$, and that the central and satellite occupations are independent, $\langle N_cN_s\rangle=\langle N_c\rangle\langle N_s\rangle$. Finally, the halo model assigns the mean SFR to all the haloes of a given mass. To account for the scatter around this mean, as in other models of the extragalactic CO emission~\citep[e.g.][]{Kokron:2024}, we add a mean-preserving lognormal scatter with $\sigma_{\rm SFR}=0.26$ dex. The sum of the three terms, $P_{{\rm CO}(J)\times {\rm CO}(J^\prime)}^{\rm tot} = P_{{\rm CO}(J)\times {\rm CO}(J^\prime)}^{\rm 1h} + P_{{\rm CO}(J)\times {\rm CO}(J^\prime)}^{\rm 2h} + P_{{\rm CO}(J)\times {\rm CO}(J^\prime)}^{\rm shot}$, enters the integral of Eq.~\eqref{eq:clCO_single}. The total CO power spectrum is obtained by summing the contributions of all the pairs of transitions,
            \begin{equation}\label{eq:clCO_tot}
                C_{\ell,\nu\times\nu^\prime}^{{\rm CO}\times {\rm CO}} = \sum_{J=1}^{J_{\rm max}}\sum_{J^\prime=1}^{J_{\rm max}}C_{\ell,\nu\times\nu^\prime}^{{\rm CO}(J)\times {\rm CO}(J^\prime)} ,
            \end{equation}
            with $J_{\rm max}=13$ in our baseline model. Only pairs of transitions whose redshift ranges overlap in the two channels contribute to the sum.

        \subsection{\cocib\,cross spectrum}\label{subsec:cocib}
            The same formalism can be extended to the \cocib\,cross-spectrum. The contribution of the transition $J$ is
            \begin{align}\label{eq:clCOCIB_single}
                C_{\ell,\nu\times\nu^\prime}^{{\rm CO}(J)\times {\rm CIB}} &= \int{\rm d}z\,\dfrac{{\rm d}\chi}{{\rm d}z}\dfrac{1}{\chi^2}\,\xi_{{\rm CIB},\nu}(z)\,\xi^J_{\nu^\prime}(z)\\
                &\times P_{{\rm CO}(J)\times {\rm CIB}}\!\left(k=\dfrac{\ell+1/2}{\chi},z\right) ,
            \end{align}
            where
            \begin{itemize}
                \item $\xi_\nu^J(z)$ is the CO emission kernel of Eq.~\eqref{eq:emiss_kernel};
                \item $\xi_{\rm CIB,\nu}(z)$ is the emission kernel of the CIB,
                \begin{equation}\label{eq:cib_kernel}
                    \xi_{{\rm CIB},\nu}(z) = \dfrac{c}{4\pi}\dfrac{1}{H(z)\,(1+z)^2\,\Delta\nu}\,\overline{j}_{{\rm CIB},\nu}(z) ,
                \end{equation}
                with $\overline{j}_{\rm CIB,\nu}(z)$ the infrared luminosity density emitted at redshift $z$. To capture the correlation between CO and CIB emission consistently, we model the infrared luminosity of galaxies with the~\cite{Kennicutt:1998} relation, $L_{\rm IR} = \kappa_{\rm IR}\rm{SFR}$, starting from the same SFR used for the CO emission, and assign it a modified black-body spectral energy distribution (SED). We checked that the resulting CIB emission is consistent with that obtained with other prescriptions for the infrared luminosity and the SED. We use $\kappa_{\rm IR}=10^{10}L_\odot/(M_\odot \rm{yr}^{-1})$ and a modified black body with $T_d=34$K and $\beta=1.6$ (see Appendix~\ref{app:kernels})\footnote{Including the SFR scatter of Sect.~\ref{subsec:coco}, the CIB power at $\ell=3000$ is $D_\ell=12.20$\muk\,at 150\ghz\, and $109.7$\muk\,at 220\ghz, to be compared with the $11.3\pm0.6$\muk\,measured by ACT DR6 at 150\ghz\,~\citep{Louis:2025} and the 11.3 and 121\muk\,measured by SPT at 150 and 220\ghz, respectively~\citep{Reichardt:2021}};
                \item $P_{\rm CO(J)\times\rm CIB}(k,z)$ is the luminosity-weighted \cocib\,power spectrum computed in the halo model.
            \end{itemize}
            The computation of $P_{\rm CO(J)\times\rm CIB}(k,z)$ follows the same steps as in Sect.~\ref{subsec:coco}, with the two fields
            \begin{equation}
                \begin{aligned}
                     \hat W_u(M,k)&\equiv \hat W_{\rm CIB} =\dfrac{\langle L_c^{\rm CIB}\rangle+\langle L_s^{\rm CIB}\rangle\,\tilde{u}_{\rm gal}(k,M)}{\overline{j}_{\rm CIB}},\\
                     \hat W_v(M,k)&\equiv \hat W_{{\rm CO}(J)}=\dfrac{\langle L_c^{J}\rangle+\langle L_s^{J}\rangle\,\tilde{u}_{\rm gal}(k,M)}{\overline{j}_{J}} .
                \end{aligned}
            \end{equation}
            The 1h, 2h, and shot noise terms are
            \begin{equation}\label{eq:P1h_cocib}
                \begin{aligned}
                    P_{{\rm CO}(J)\times {\rm CIB}}^{\rm 1h} &= \dfrac{1}{\overline{j}_J\,\overline{j}_{\rm CIB}}\int {\rm d}M\,\dfrac{{\rm d}n}{{\rm d}M}\biggl[\Bigl(N_c^{\rm CIB}N_s^{\rm CO}\,\ell_c^{\rm CIB}\ell_s^{J}\\
                    &+N_s^{\rm CIB}N_c^{\rm CO}\,\ell_s^{\rm CIB}\ell_c^{J}\Bigr)\tilde{u}_{\rm gal} + N_s^{\rm CIB}N_s^{\rm CO}\,\ell_s^{\rm CIB}\ell_s^{J}\,\tilde{u}_{\rm gal}^2\biggr] ,
                \end{aligned}
            \end{equation}
            \begin{equation}\label{eq:P2h_cocib}
                \begin{aligned}
                    P_{{\rm CO}(J)\times{\rm CIB}}^{\rm 2h} &=P_{mm}^{\rm lin}(k)\biggl[\dfrac{1}{\overline{j}_{J}}\int {\rm d}M\,\dfrac{{\rm d}n}{{\rm d}M}\,b(M)\Bigl(N_c^{\rm CO}\ell_c^{J}+N_s^{\rm CO}\ell_s^{J}\tilde{u}_{\rm gal}\Bigr)\biggr]\\
                    &\times\biggl[\dfrac{1}{\overline{j}_{\rm CIB}}\int {\rm d}M\,\dfrac{{\rm d}n}{{\rm d}M}\,b(M)\Bigl(N_c^{\rm CIB}\ell_c^{\rm CIB}+N_s^{\rm CIB}\ell_s^{\rm CIB}\tilde{u}_{\rm gal}\Bigr)\biggr] ,
                \end{aligned}
            \end{equation}
            \begin{equation}\label{eq:Pshot_cocib}
                \begin{aligned}
                    P_{{\rm CO}(J)\times{\rm CIB}}^{\rm shot} &=  \dfrac{1}{\overline{j}_J\,\overline{j}_{\rm CIB}}\int {\rm d}M\,\dfrac{{\rm d}n}{{\rm d}M}\biggl[N_c^{CO\cap CIB}\,\ell_c^{J}\ell_c^{\rm CIB}
                    \\
                    &+N_s^{CO\cap CIB}\,\ell_s^{J}\ell_s^{\rm CIB}\biggr] ,
                \end{aligned}
            \end{equation}
            respectively, where $\ell_{\rm c/s}^{\rm CIB}$ is the infrared luminosity of a central/satellite galaxy in the channel $\nu$, $\ell^{\rm CIB}_{c/s,\nu}=\kappa_{\rm IR}\,{\rm SFR}_{c/s}\,f_\nu(z)$, with $f_\nu(z)$ the fraction of the SED emitted in the rest-frame band of the channel. The shot noise term receives contributions only from galaxies that emit in both CO and the infrared, since it describes the correlation of each galaxy with itself; $N^{CO\cap CIB}_{c/s}$ is therefore the mean number of central/satellite galaxies per halo belonging to both populations. The two HODs fix $N^{\rm CO}_{c/s}$ and $N^{\rm CIB}_{c/s}$ separately. We assume $N^{CO \cap CIB}_{c/s}=\min\bigl(N^{\rm CO}_{c/s},N^{\rm CIB}_{c/s}\bigr)$. The 1h term, instead, involves pairs of distinct galaxies: for Poisson-distributed satellites their number is $N_s^{\rm CO}N_s^{\rm CIB}$ regardless the overlap between the two populations, so Eq.~\eqref{eq:P1h_cocib} does not depend on $N^{CO \cap CIB}$. The full derivation is given in Appendix~\ref{app:hm_calc}.

            As for the CO auto-spectrum, the total \cocib\,cross-spectrum is obtained by summing over all the transitions,
            \begin{equation}\label{eq:clCOCIB_tot}
                C_{\ell,\nu\times\nu^\prime}^{{\rm CO}\times {\rm CIB}} = \sum_{J=1}^{J_{\rm max}}C_{\ell,\nu\times\nu^\prime}^{{\rm CO}(J)\times {\rm CIB}} .
            \end{equation}
            Since the CIB is a continuum, every CO transition that falls in the channel $\nu^\prime$ correlates with the CIB emitted at the same redshift in any channel $\nu$. Note that $C_{\ell,\nu\times\nu^\prime}^{{\rm CO}\times{\rm CIB}}\neq C_{\ell,\nu^\prime\times\nu}^{{\rm CO}\times{\rm CIB}}$ for $\nu\neq\nu^\prime$. The \cocib\,contribution to the total sky power spectrum between $\nu$ and $\nu^\prime$ is the sum of the two orderings, $C_{\ell,\nu\times\nu^\prime}^{{\rm CO}\times{\rm CIB}}+C_{\ell,\nu^\prime\times\nu}^{{\rm CO}\times{\rm CIB}}$.

\section{Model assumptions}\label{sec:mod_ass}
    In this section, we review the main assumptions of the model and quantify their impact on the power spectra. Section~\ref{subsec:co_lum} investigates the modelling of the CO line luminosity, comparing different CO SLEDs in Sect.~\ref{subsec:co_sled} and different SFR prescriptions in Sect.~\ref{subsec:co_sfrs}. Sect.~\ref{subsec:co_ranges} discusses the numerical choices for the halo mass range and the maximum CO transition included in the calculation. All figures in this section show the \coco\,power spectrum for the \lact\,channels, with the instrument specifications listed in Table~\ref{tab:act_features}.

    Compared with previous predictions of CO in CMB bands~\citep{Maniyar:2023, Kokron:2024, Mehta:2025}, the main features of our approach are that the \coco\,and the \cocib\,spectra are computed analytically in the halo model, with independent HODs for the two tracers and that the CO and infrared luminosities derive from the same SFR, so that the \cocib\,correlation is a prediction of the model. In addition to that, our pipeline includes all the transitions up to $J=13$, and both central and satellite galaxies. We compare our predictions with those of previous works in Sect.~\ref{subsec:discussion_cmbsky}.

    \subsection{CO line luminosity}\label{subsec:co_lum}
        We model the CO luminosity of a galaxy through a direct link with its SFR,
        \begin{equation}\label{eq:lco}
            L_{\rm CO}^{J} = R_{J}\,{\rm SFR}(M,z) .
        \end{equation}
            The coefficients $R_J$ are the weights assigned to each transition of the CO ladder, defined through the CO SLED. Their normalisation is set by the $\rm{CO}(1\rightarrow0)$ coefficient, $R_1$, which we relate to the molecular gas depletion time, $t_{\rm dep}\equiv M_{\rm H_2}/\rm{SFR}$. The molecular gas mass is obtained from the $\rm{CO}(1\rightarrow0)$ line luminosity through the CO-to-H$_2$ conversion factor $\alpha_{\rm CO}$, $M_{\rm H_2}=\alpha_{\rm CO}L_{\rm CO(1\rightarrow0)}^\prime$, where $L^\prime$ is the line luminosity in $\rm{K\,km\,s^{-1}pc^2}$. The latter is related to the luminosity in $L_\odot$ by $L=3.2\times10^{-11}(\nu_{\rm rest}/\rm{GHz})^3L^\prime$~\citep{Solomon:2005}. Therefore,
        \begin{equation}\label{eq:r1}
            R_1=\dfrac{L_{\rm CO(1\rightarrow0)}}{\rm{SFR}} = 3.2\times10^{-11}\biggl(\dfrac{\nu_{\rm rest}}{\rm{GHz}}\biggr)^3\dfrac{t_{\rm dep}}{\alpha_{\rm CO}} .
        \end{equation}
        We adopt $t_{\rm dep}=0.7$\Gyr\,and $\alpha_{\rm CO}=3.6 M_\odot(\rm{K\,km\,s^{-1}pc^2})^{-1}$, typical of main-sequence galaxies~\citep{Tacconi:2018, Tacconi:2020, Decarli:2020}. In our baseline model, the SFR is taken from the smooth \um\,relation (B19, described in Sect.~\ref{subsec:co_sfrs}). In this section, we evaluate how different prescriptions for both the CO SLED and the SFR affect the power spectrum.

        \subsubsection{CO spectral line energy distribution}\label{subsec:co_sled}
\begin{figure}[h!]
                \centering
                \includegraphics[width=0.8\linewidth]{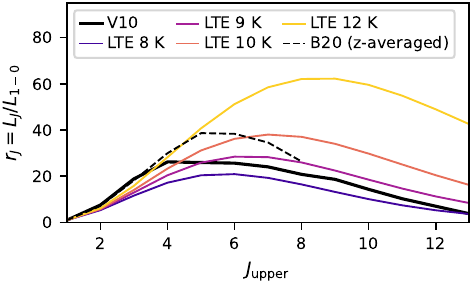}
                \caption{CO SLEDs normalised to CO(1$\rightarrow$0): the empirical V10 SLED (solid black) and the LTE SLED of Eq.~\eqref{eq:lte_sled} for $T_{\rm ex}=8$--$12$\,K (coloured lines). The dashed curve shows the B20 SLED prescription and stops at $J=8$ as it is the highest $J$-transition included in B20 analysis.}
                \label{fig:sled_comparison}
            \end{figure}

            The coefficients $R_J$ of Eq.~\eqref{eq:lco} encode the CO SLED and assign different weights to the different transitions. We consider three main prescriptions. Our baseline model uses the shape of the empirical conversion coefficients of \citet[][V10 hereafter]{Visbal2010}, reported in their Table 1. These coefficients are calibrated on the starburst galaxy M82 and imply $t_{\rm dep}=0.27$\Gyr. In our baseline, we keep their ratios $R_J/R_1$ and multiply all $R_J$ by 2.57, so that $R_1$ satisfies Eq.~\eqref{eq:r1}: with $t_{\rm dep}=0.7$\Gyr\,and $\alpha_{\rm CO}=3.6$, Eq.~\eqref{eq:r1} gives $R_1\simeq9.5\times10^3L_\odot/(M_\odot\rm{yr}^{-1})$, compared with $3.7\times10^3$ for V10. The 2.57 factor is not a free parameter but the ratio of the two normalisations: it moves the normalisation from the starburst M82 to main-sequence galaxies while keeping the V10 SLED shape. As an alternative, we consider an analytic SLED corresponding to a single-temperature population in local thermodynamic equilibrium (LTE) with excitation temperature $T_{\rm ex}$. In this case, the luminosity ratios relative to $\rm{CO}(1\rightarrow0)$ are
            \begin{equation}\label{eq:lte_sled}
                r_J(T_{\rm ex})\equiv\dfrac{L_J}{L_1} = J^4\,\dfrac{e^{T_1/T_{\rm ex}}-1}{e^{JT_1/T_{\rm ex}}-1} ,
            \end{equation}
            with $T_1=h\nu_1/k_B\simeq 5.5 K$, and $R_J=r_JR_1$. Low excitation temperatures, $T_{\rm ex}\sim8-12 K$, describe the sub-thermally excited gas expected in typical star-forming galaxies. Figure~\ref{fig:sled_comparison} compares the V10 SLED with the LTE SLEDs for $T_{\rm ex}\in[8,12]K$. The V10 SLED peaks at $J\simeq4-6$ and falls off faster at high $J$ than the LTE models, which peak at $J\simeq6-9$ depending on $T_{\rm ex}$. Finally, we consider the average CO SLED of main-sequence galaxies measured by ASPECS~\citep[][B20 hereafter]{Boogaard:2020}.

\begin{figure*}[h!]
                \centering
                \includegraphics[width=0.9\linewidth]{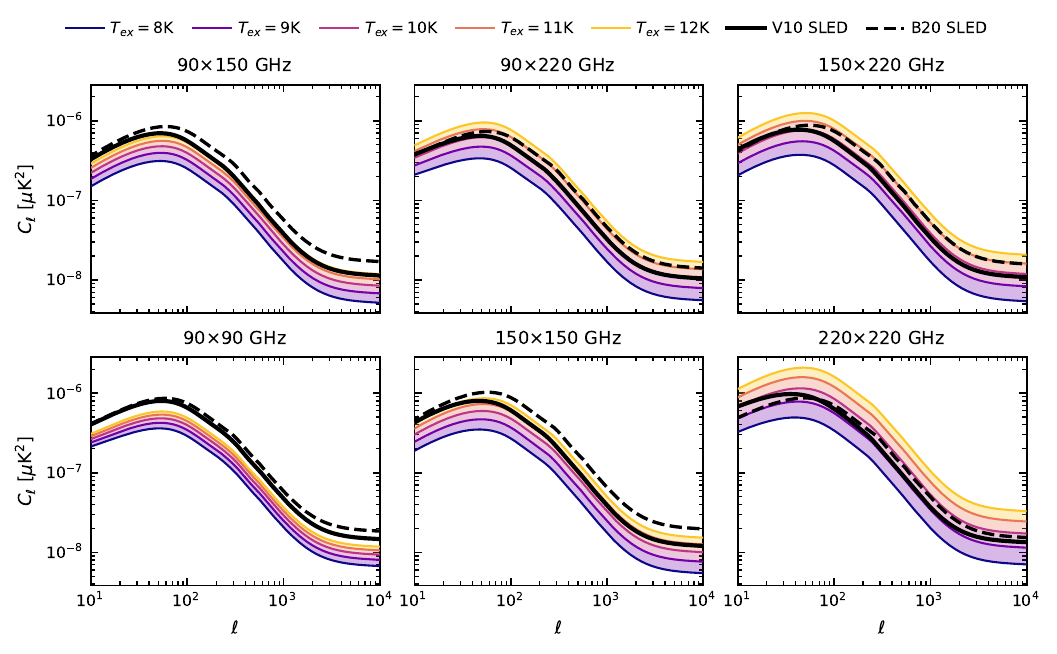}
                \caption{\coco\,angular power spectra for the \lact\,auto- and cross-frequency channels, for the V10 SLED (solid black), for the LTE SLED with $T_{\rm ex}$ between 8 and 12\,K (coloured lines and shaded region), and for the B20 SLED (dashed black).}
                \label{fig:sled_ps_comparison}
            \end{figure*}

            Figure~\ref{fig:sled_ps_comparison} shows the effect of the three prescriptions on the \coco\,power spectrum. Varying $T_{\rm ex}$ between 8 and 12 K changes the amplitude of the spectra by a factor of $\sim 2$ at 90\ghz\,and up to $\sim 5$ at 220\ghz. The V10 SLED at 90\ghz\,is above the LTE curves because of the larger $\rm{CO}(2\rightarrow1)$ weight we can see in Fig.~\ref{fig:sled_comparison}. The \cocib\,cross-spectrum is less sensitive to the SLED, which enters only one of the two fields.

            The analytic prescription introduces an additional free parameter, $T_{\rm ex}$. As shown in Sect.~\ref{sec:results}, even a \lcmnhd\,experiment has limited sensitivity to the parameters describing CO emitters. We therefore do not expect CMB data alone to constrain $T_{\rm ex}$, and we adopt the empirical V10 prescription in our baseline model as the simplest choice that is defined up to $J=13$; the B20 SLED is measured only up to $J=8$. All the implementations are available in the code.

        \subsubsection{SFR prescriptions}\label{subsec:co_sfrs}
\begin{figure}
                \centering
                \includegraphics[width=0.8\linewidth]{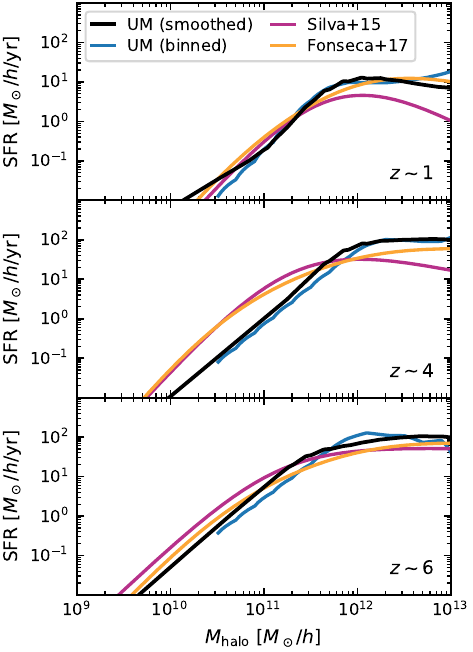}
                \caption{SFR as a function of halo mass at $z\sim1$, 4, and 6 for the B19 baseline (black), for the analytic prescriptions of Silva+15 and Fonseca+17, and for the binned \um\,SFR.}
                \label{fig:sfrs}
            \end{figure}

            Our baseline SFR (B19 hereafter) is the smooth $\rm{SFR}(M,z)$ of \um\,as tabulated in the \texttt{LIMpy} package\footnote{\href{https://github.com/Anirbancosmo/Limpy.git}{\color{magenta}\faGithub\,\, Anirbancosmo/Limpy}} \citep{Roy:2023limpy}. In addition, we consider the \um\,SFR of central and satellite galaxies binned in halo mass (\um\,binned), and two analytic SFR--halo mass relations commonly used in LIM studies and implemented in the \texttt{LIMpy} package as well:
            \begin{itemize}
                \item Silva+15, the $\rm{SFR}(M,z)$ fitting function of \cite{Silva:2015}, calibrated on simulated galaxy catalogues;
                \item Fonseca+17, the $\rm{SFR}(M,z)$ parametrisation of \cite{Fonseca:2017}.
            \end{itemize}
            Figure~\ref{fig:sfrs} compares these relations at $z\sim1,\,4,$ and $6$, showing for all of them the total SFR of the halo, i.e. central plus satellite galaxies.

\begin{figure*}
                \centering
                \includegraphics[width=0.9\linewidth]{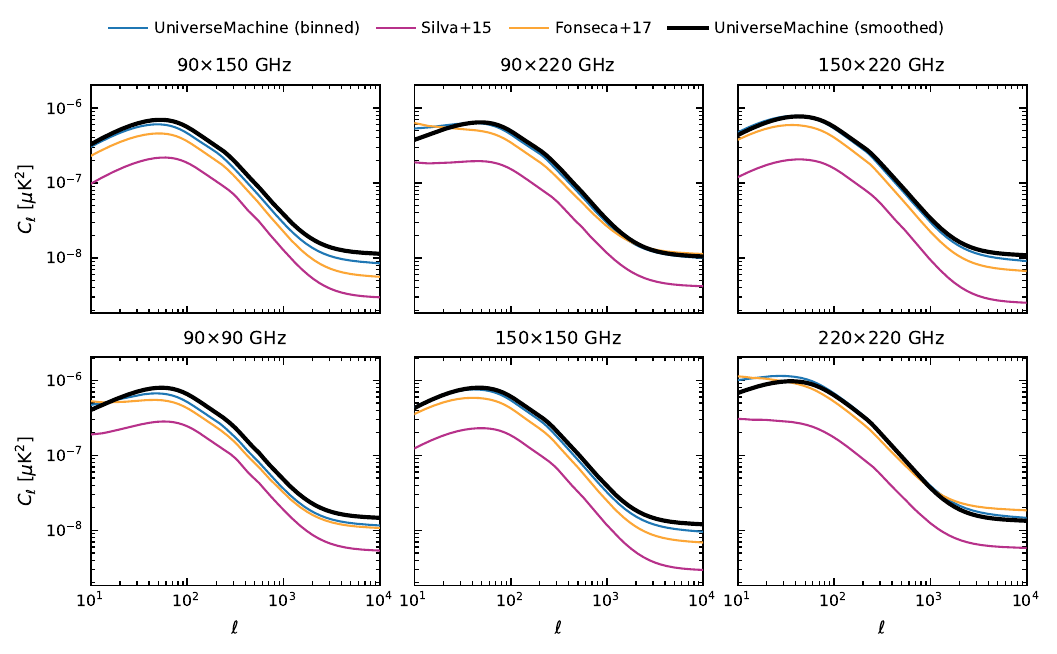}
                \caption{\coco\,angular power spectra for the \lact\,auto- and cross-frequency channels, for the B19 baseline SFR (solid black) and for the Silva+15, Fonseca+17, and binned \um\,prescriptions (coloured lines).}
                \label{fig:sfr_ps_comparison}
            \end{figure*}

            Figure~\ref{fig:sfr_ps_comparison} shows the effect of the SFR model on the \coco\,power spectra. We show the \lact\,auto- and cross-frequency spectra. The four SFR models considered in the analysis show similar shape, while their amplitudes span a factor of $\sim 3-4$. Our baseline B19 model gives the highest power at 90 and 150\ghz, and lies within the spread at 220\ghz. The SFR model is thus, together with the SLED, the main source of uncertainty in the amplitude of the CO power spectrum. This agrees with \cite{Mehta:2025}, who found an even larger spread when varying the SFR--$L_{\rm CO}$ relation. The \cocib\,spectra show the same behaviour, with a smaller spread.

            The SFR prescriptions implemented in \texttt{LIMpy} give a single SFR per halo. For a consistent comparison, we treat all of them as the total SFR of the halo and split it between central and satellite galaxies using the satellite fraction of \um
            \,(see Sect.~\ref{subsec:coco}). Appendix~\ref{app:co_gal} quantifies the relative contribution of satellite galaxies to the \coco\,and the \cocib\,power spectra: at $\ell\gsim2000$, they increase the power by $\simeq45\%$ and $\simeq25\%$, respectively.

    \subsection{Halo mass range and maximum transition}\label{subsec:co_ranges}
        We also evaluate the impact of the halo mass range used in the integrals and of the maximum transition $J_{\rm max}$ included in the sums of Eqs.~\eqref{eq:clCO_tot} and~\eqref{eq:clCOCIB_tot}.

        For the halo mass range, we start from a wide range, $M\in[10^9, 10^{15}]M_{\odot}$, and compute the power spectrum for different values of the minimum halo mass, $M_{\rm low}=10^9,\,10^{9.5},\,10^{10},\,10^{10.5},\,10^{11},\,10^{11.5}\,M_{\odot}$. Our analysis shows that considering minimum masses smaller than $10^{10.5}M_\odot$ affects neither the \coco\,nor the \cocib\,power spectra. Therefore, our baseline uses $M_{\rm low}=10^{10.5}M_\odot h^{-1}$.

\begin{figure}
            \centering
            \includegraphics[width=\linewidth]{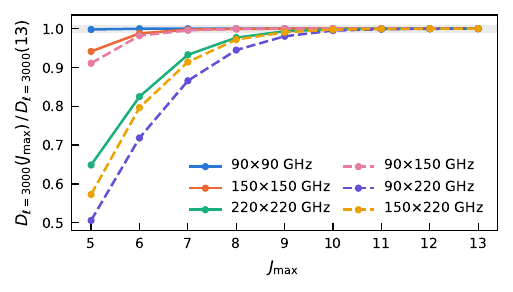}
            \caption{Ratio of the \coco\,power spectra to the baseline case ($J_{\rm max}=13$) evaluated at $\ell=3000$. Solid curves represent auto-frequency spectra, and dashed curves cross-frequency spectra.}
            \label{fig:Jmax_ps}
        \end{figure}

\begin{figure}
            \centering
            \includegraphics[width=0.7\linewidth]{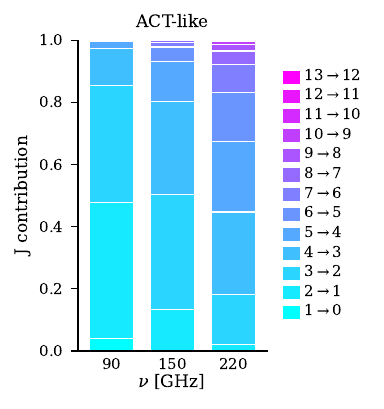}
            \caption{Relative contribution of each CO transition to the band-averaged mean CO intensity in the \lact\,channels, computed for the baseline model (B19 and V10 SLED), and the top-hat bandpasses of Table~\ref{tab:act_features}. The colours of the transitions correspond to those of Fig.~\ref{fig:line_int}.}
            \label{fig:J_chan}
        \end{figure}

        Figure~\ref{fig:Jmax_ps} shows the effect of truncating the CO ladder at $J_{\rm max}<13$. The 90\ghz\,auto-spectrum is insensitive to $J_{\rm max}$ down to $J_{\rm max}=5$, because transitions with $J>5$ do not fall in the 90\ghz\,bandwidth. For the same reason, the sensitivity increases with frequency, since higher $J$-transitions contribute more to the highest channels. Transitions with $J\geq10$ contribute less than $2\%$ of the \coco power at $\ell=3000$ in all channels\footnote{And less than $4\%$ of the \cocib\,power at $\ell=3000$.}. Figure~\ref{fig:J_chan} shows the relative contribution of each \textit{J}-transition to the channels considered in this analysis, based on Eq.~\eqref{eq:zrange}.

\section{Fisher forecasts}\label{sec:results}
    In this section, we assess the constraining power of current and future CMB experiments on the model parameters with Fisher forecasts for \lact\,and \lcmnhd\,configurations, and we quantify the bias induced on cosmological and foreground parameters when CO is present in the data but neglected in the model (Sect.~\ref{subsec:noco}). The two configurations bracket the current and the next generations of high-resolution experiments: the \lact\,configuration is representative of current data, while the \lcmnhd\,configuration is the most sensitive experiment proposed so far. Experiments of intermediate sensitivity, such as the Simons Observatory~\citep{SimonsObservatory:2018koc}, are expected to give results between the two cases, and are left for future work.

    To this end, we use the Fisher formalism~\citep[e.g.][]{Tegmark:1997}, a standard tool to forecast how well a given dataset can constrain the parameters of a model, without the need for simulated data realisations. For a set of $N$ model parameters $\{\theta_i\}$, the Fisher matrix $\mathbf{F}$ is an $N\times N$ symmetric matrix,
    \begin{equation}\label{eq:fisher}
        \mathbf{F} = \mathbf{D}\mathbf{\Sigma}^{-1}\mathbf{D}^T,
    \end{equation}
    where $\mathbf{\Sigma}$ is the covariance matrix of the data vector, which contains the binned power spectra of all the independent auto- and cross-frequency combinations, and $\mathbf{D}$ is the matrix of the derivatives of the data vector with respect to the parameters. We use the Gaussian covariance of the power spectra,
    \begin{equation}\label{eq:covmat}
        \Sigma\!\left(C_\ell^{\nu_1\nu_2}, C_{\ell^\prime}^{\nu_3\nu_4}\right)
        = \frac{\delta_{\ell\ell^\prime}}{(2\ell+1)\,f_{\rm sky}\,\Delta\ell}
        \left(C_\ell^{\nu_1\nu_3}C_\ell^{\nu_2\nu_4}
             + C_\ell^{\nu_1\nu_4}C_\ell^{\nu_2\nu_3}\right) ,
    \end{equation}
    where $f_{\rm sky}$ is the observed sky fraction, $\Delta\ell$ is the width of the multipole bins, and $C_\ell^{\nu\nu^\prime}$ are the total power spectra, i.e. signal plus the instrumental noise. The signal contains the primary CMB and the foregrounds. The CMB power spectrum is computed with \camb\,\citep{Lewis:1999bs, Howlett:2012mh} for a \planck 18 cosmology~\citep{Planck:2018vyg} with the values of the cosmological parameters listed in Table~\ref{tab:fg_params}. The foregrounds include the usual extragalactic components of CMB analyses (CIB, tSZ, tSZ$\times$CIB, kSZ, and radio sources) and the new components introduced in this work, \coco\, and \cocib. CIB, tSZ, and tSZ$\times$CIB are described in a halo model framework consistent with the CO model, the kSZ uses the template of~\cite{George:2014oba}, and radio sources are modelled as a Poisson term. The foreground model is described in Appendix~\ref{app:fg_mod}. We assume white instrumental noise and Gaussian beams, with the map noise levels and  and beam widths listed in Table~\ref{tab:act_features}. For the \lcmnhd\,configuration, we complement the Gaussian covariance with a systematic error floor, as discussed later in this section.

    The marginalised uncertainty on the parameter $\theta_i$ is $\sigma_i=\sqrt{(\mathbf{F}^{-1})_{ii}}$. We quantify the constraining power through the signal-to-noise ratio $\rm{SNR}=\theta_{i,0}/\sigma_i$ and the degeneracies through the correlation matrix $\rho_{ij}=(\mathbf{F}^{-1})_{ij}/(\sigma_i\sigma_j)$. The parameter set consists of six cosmological parameters, $\{\mathcal{A}_s,\Omega_m,\Omega_b,\tau,n_s,h\}$, the amplitude and the spectral index of radio sources, $\{\mathcal{A}_{\rm radio},\alpha_{\rm radio}\}$, the kSZ amplitude $\mathcal{A}_{\rm kSZ}$, the hydrostatic mass bias $b_{HSE}$, which sets the amplitude of the tSZ signal, and the HOD parameters of CO and CIB emitters defined in Sect.~\ref{sec:model}. For each experimental configuration, we consider two main cases:
    \begin{itemize}
        \item[A.] cosmological and foreground parameters varied consistently in the CMB and in the foreground model; CO emitted by central and satellite galaxies (18 parameters);
        \item[B.] cosmological and foreground parameters varied consistently in the CMB and in the foreground model;     CO emitted by central galaxies only (16 parameters).
    \end{itemize}
    Case A is our baseline.

    \subsection{Parameter constraints}
        Here, we present the constraints on the model parameters forecast for the \lact\,and \lcmnhd\,configurations with the formalism described above. For each configuration, we show the correlation matrix and the SNR of the parameters for our baseline, case A, and we then discuss how the results change when CO is emitted by central galaxies only (case B). The marginalised uncertainties are reported in Appendix~\ref{app:fisher} (Table~\ref{tab:fisher_errors}).

        \subsubsection{Fisher analysis: \lact}\label{subsub:actL}
        For the \lact\, configuration, we consider the three nominal ACT frequency channels (90, 150, and 220\ghz), the multipole range $\ell\in[1000,8500]$, a sky fraction of $f_{\rm sky}=0.23$, which corresponds to the $\simeq9400\,\rm{deg}^2$ of ACT DR6 data left after masking~\citep{ACT:2023dou} and is comparable with the $\simeq25\%$ of the sky used in the DR6 power spectrum analysis of~\cite{Louis:2025}, and a linear binning with $\Delta\ell=50$.
        \begin{table}
            \centering
            \caption{Specifications of the \lact\ and \lcmnhd\ configurations.}
            \label{tab:act_features}
            \begin{tabular}{ccccc}
               \hline\hline
               $\nu_{\rm central}$ & $\nu_{\rm min}$ & $\nu_{\rm max}$ & $\sigma_T$ & $\theta_{\rm FWHM}$\\
               $[{\rm GHz}]$ & $[{\rm GHz}]$ & $[{\rm GHz}]$ & $[\mu{\rm K\,arcmin}]$ & $[{\rm arcmin}]$ \\
               \hline
               \multicolumn{5}{c}{\textit{\lact}}\\
               \hline
               90  & 77  & 112 & 15.09 & 2.0\\
               150 & 124 & 172 & 18.22 & 1.4\\
               220 & 182 & 277 & 82.00 & 1.0\\
               \hline
               \multicolumn{5}{c}{\textit{\lcmnhd}}\\
               \hline
               30  & 25.5  & 34.5  & 6.5  & 1.25\\
               40  & 34    & 46    & 3.4  & 0.94\\
               90  & 76.5  & 103.5 & 0.73 & 0.42\\
               150 & 127.5 & 172.5 & 0.79 & 0.25\\
               220 & 187   & 253   & 2.0  & 0.17\\
               280 & 238   & 322   & 2.7  & 0.13\\
               \hline
            \end{tabular}
        \end{table}

        \begin{figure*}
            \centering
            \includegraphics[height=7.6cm]{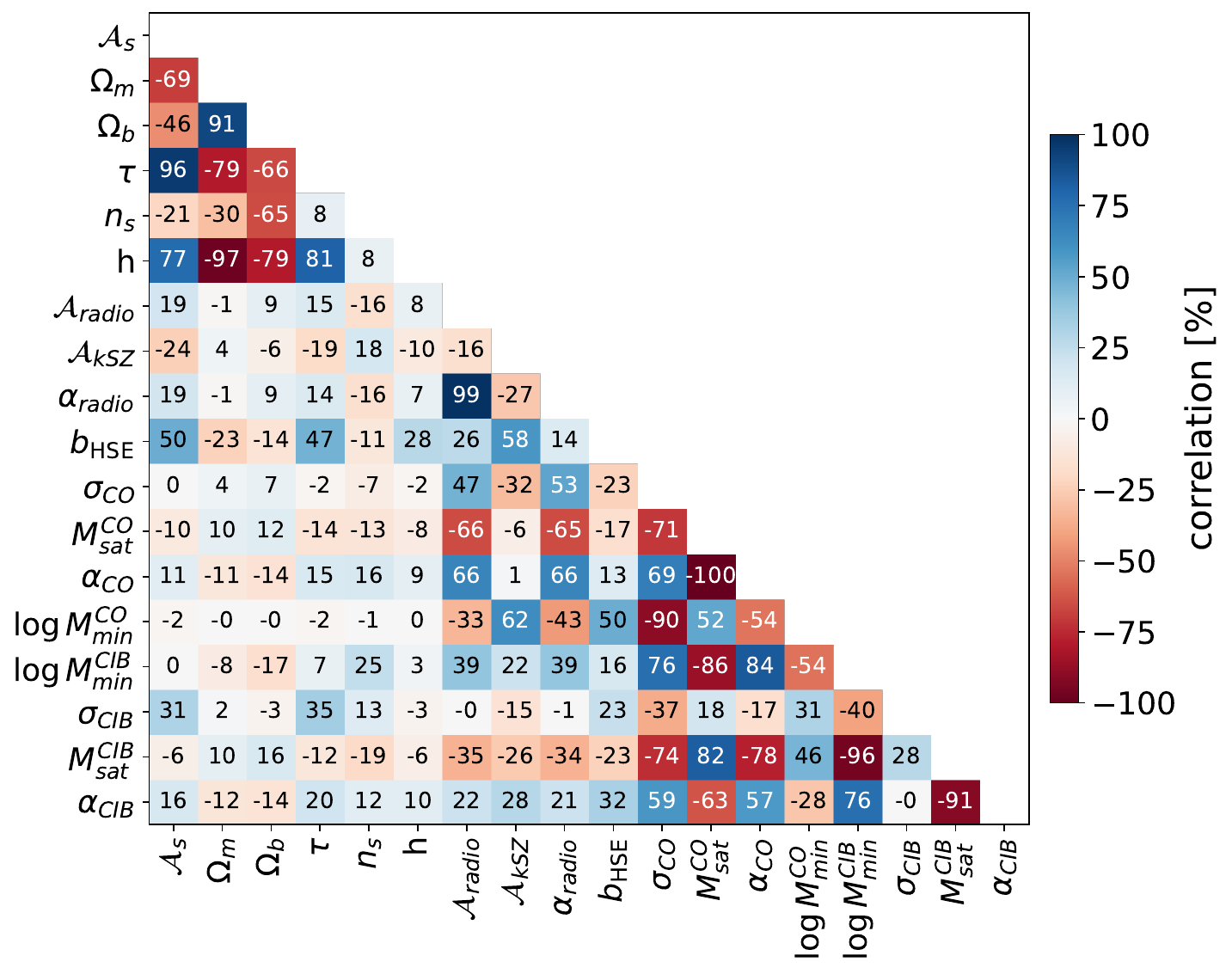}
            \includegraphics[height=7.6cm]{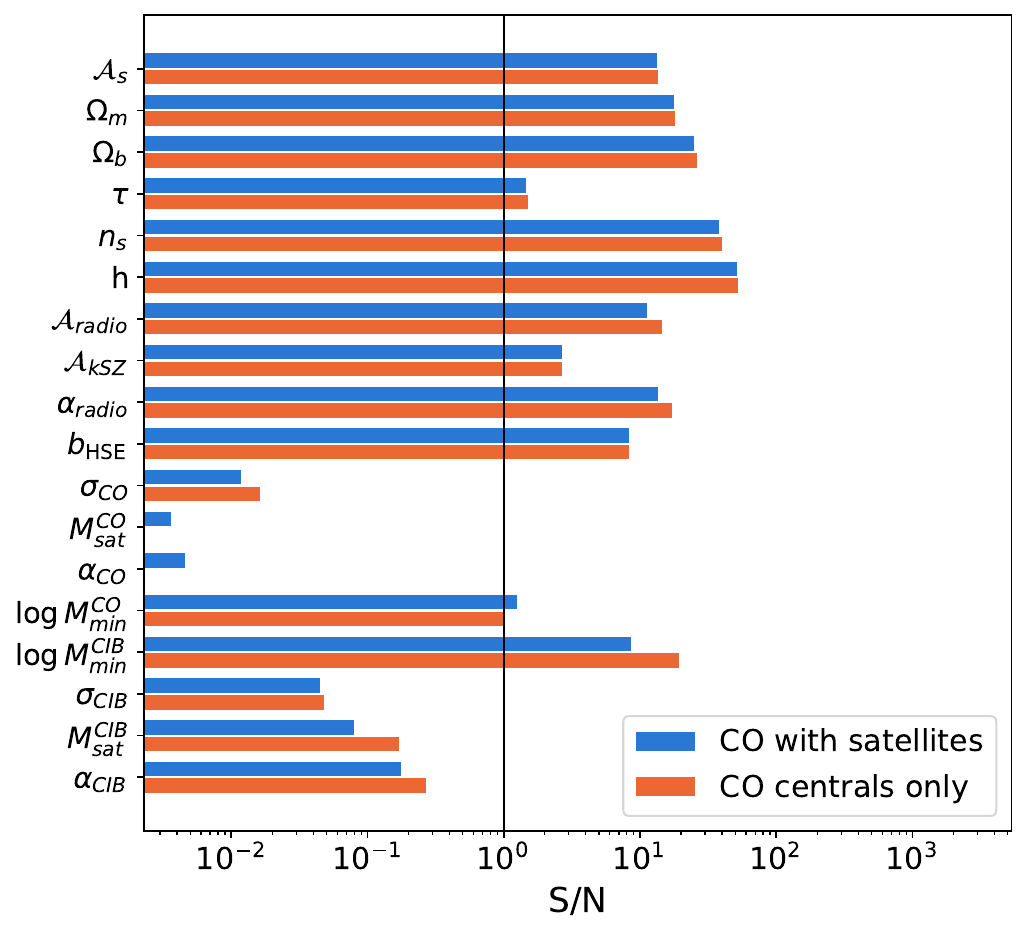}
            \caption{Fisher forecast for the \lact\,configuration. \textit{Left}: correlation matrix of the parameters (in per cent) for case A (cosmological and foreground parameters varied, CO from central and satellite galaxies). \textit{Right}: marginalised SNR, $\theta_{i,0}/\sigma_i$, for case A (blue) and case B (CO from central galaxies only, orange); the vertical line marks ${\rm SNR}=1$.}
            \label{fig:fisher_act_A}
        \end{figure*}

        Figure~\ref{fig:fisher_act_A} shows the correlation matrix and the SNR of the parameters for case A, in the left and right panels, respectively. The correlation matrix has a clear block structure. The cosmological parameters are strongly correlated among themselves, and so are the foreground parameters, while the correlations between the two blocks are mostly lower. Within the cosmological block, we recover well-established degeneracies, which serve as an internal consistency test. Within the foreground block, the amplitude and spectral index of radio sources are almost fully degenerate in all cases. The tSZ amplitude, controlled by $b_{HM}$, is correlated with the kSZ amplitude ($\rho\simeq0.58$), as is well known from ACT and SPT analyses~\citep{Dunkley:2013vu, Reichardt:2021}. The occupation parameters, which set the amplitude of the CO and CIB terms, are also correlated with the radio and SZ parameters: at $\ell\gsim3000$, all these terms are smooth functions of $\ell$, and three frequency channels are not enough to separate them, so that a change in one can be compensated by a change in others. Finally, $\sigma_{\rm CO}$ and $\log M_{\rm min}^{\rm CO}$ are strongly anti-correlated ($\rho\simeq-0.90$), because increasing either of them lowers the CO power. The satellite parameters of CO and CIB emitters are also strongly correlated with each other, so that part of the CO signal can be mimicked by the occupation of CIB emitters (see Sect.~\ref{subsec:noco}).

        Only a subset of the parameters is significantly constrained. The cosmological, radio, and SZ parameters have $\rm{SNR}>1$. Among the occupation parameters, only the minimum halo mass of CIB emitters is well constrained; for that of CO emitters, $\rm{SNR}\simeq1$, i.e. the data provide an upper limit on the CO power rather than a measurement of $M_{\rm min}^{\rm CO}$, and all the other CO and CIB occupation parameters have $\rm{SNR}<1$. Note, however, that our scenario explores the full parameter space. It is well known that two out of the four HOD parameters of the CIB ($\sigma,\,M_{\rm sat}$) cannot be constrained by the data~\citep{Planck:2011ivn} and are usually fixed to values from the literature~\citep{Planck:2011ivn, Tinker:2009jp}.

        Without CO satellites (case B, Fig.~\ref{fig:app_corr_nosat}), the two parameters of the central occupation of CO emitters, $\sigma_{\rm CO}$ and $M_{\rm min}^{\rm CO}$, become fully degenerate. The removal of the two unconstrained satellite parameters increases the SNR of the radio parameters and of the minimum halo mass of CIB emitters, while the cosmological and SZ parameters are almost unaffected (see the right panel of Fig.~\ref{fig:fisher_act_A} for the comparison).

        \subsubsection{Fisher analysis: \lcmnhd}\label{subsub:cmbhdL}
        For the \lcmnhd\,configuration, we consider six frequency channels between 30 and 280\ghz\,with the specifications listed in Table~\ref{tab:act_features} \citep{Sehgal:2019,CMBHD:2022}, the multipole range $\ell\in[1000,30000]$, and a sky fraction $f_{\rm sky}=0.6$.

        For this configuration, the Gaussian covariance of Eq.~\eqref{eq:covmat} underestimates the uncertainties on the bandpowers. Thanks to the low noise, the Gaussian uncertainty of the \lcmnhd\,bandpowers is close to the cosmic-variance limit of the total signal, and the large sky fraction lowers this limit further. The contributions not included in Eq.~\eqref{eq:covmat}, i.e. the non-Gaussian term of extragalactic point sources (trispectrum), and the calibration, beam and bandpass uncertainties, affect any experiment, including the \lact\,configuration, but their impact grows as the Gaussian uncertainty decreases, and at the level of a \lcmnhd\,experiment they cannot be neglected. The bias calculation of Sect.~\ref{subsec:noco} further motivates a more realistic error budget: with the Gaussian covariance only, neglecting CO in the \lcmnhd\,forecast would -- unrealistically -- shift all the parameters by more than $10\sigma$.

        To obtain more realistic error bars without computing trispectra and systematic contributions, we modify the \lcmnhd\,forecast in two ways:
        \begin{enumerate}
            \item we add a systematic error floor, i.e. an uncorrelated error $fC_b$ at each bandpower $C_b$;
            \item we consider a cut in the maximum multipole, $\ell_{\rm max}=8500$, i.e. the multipole range of the \lact\,configuration, in addition to the full range, $\ell_{\rm max}=30000$.
        \end{enumerate}
        The floor should be understood as a simple proxy for non-Gaussian and calibration uncertainties. Since the bandpowers at $\ell\gsim3000$ are dominated by extragalactic foregrounds, whose modelling is currently accurate at the level of a few per cent, we adopt $f=3\%$ as our reference error budget, and show the results for the $f=1\%$ and $f=10\%$ cases in Appendix~\ref{app:fisher}, Fig.~\ref{fig:hd_bias_1pc}. All the \lcmnhd\,results below are obtained with this covariance.

        \begin{figure*}
            \centering
            \includegraphics[height=7.6cm]{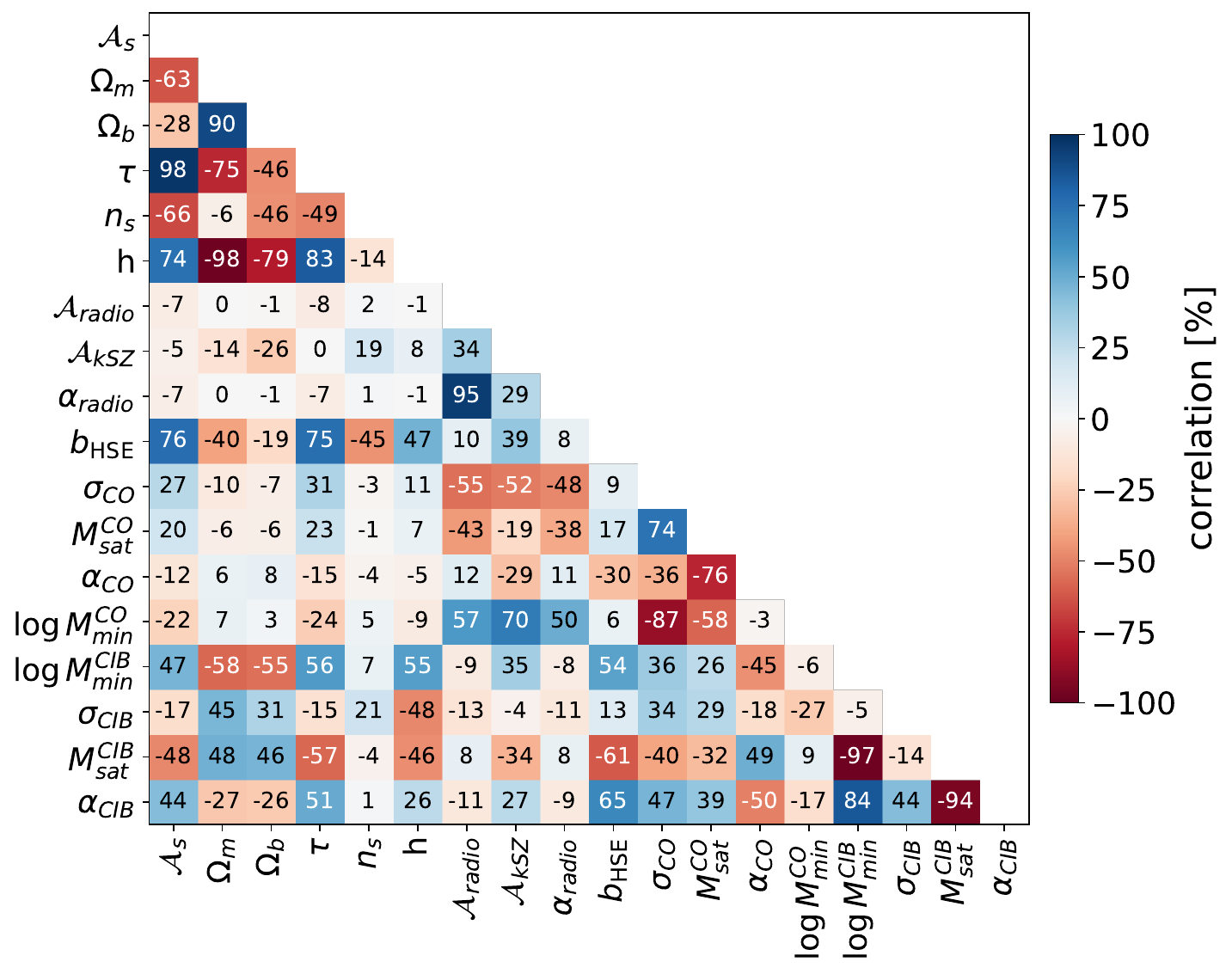}\hfill
            \includegraphics[height=7.6cm]{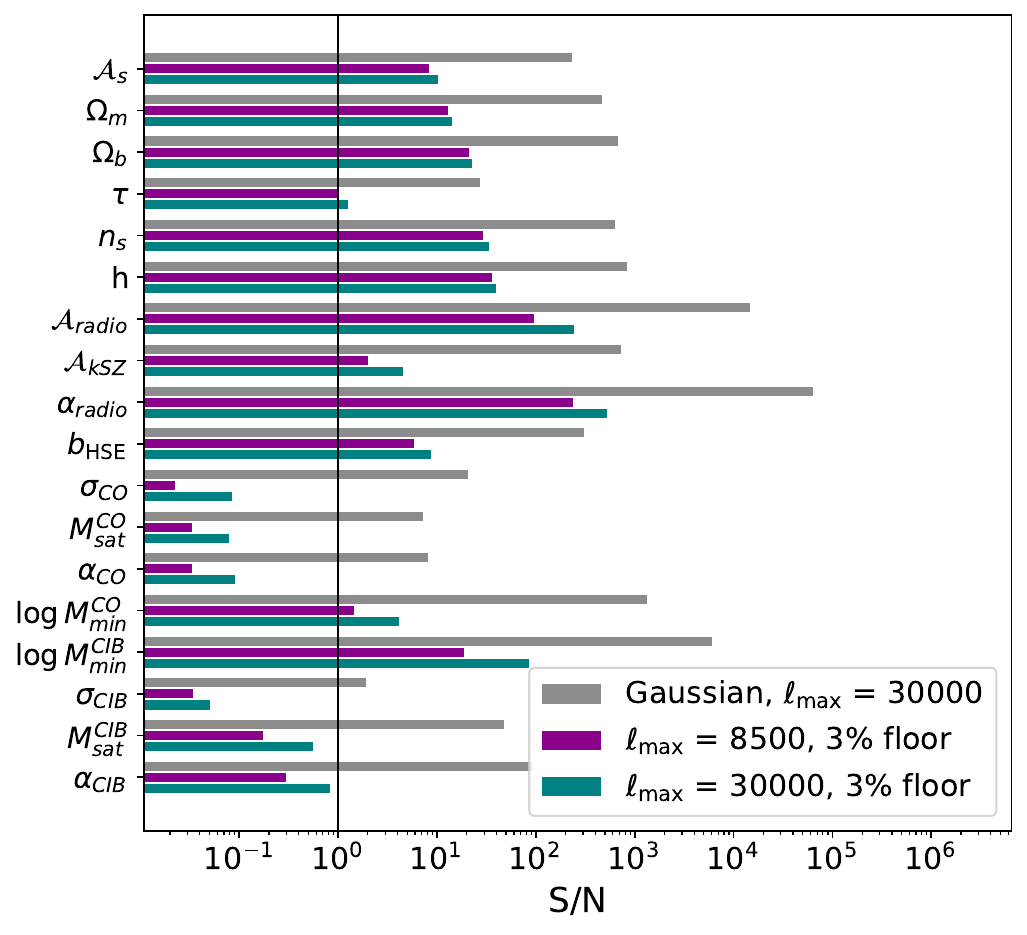}
            \caption{Fisher forecast for the \lcmnhd\,configuration (case A). \textit{Left}: correlation matrix of the parameters with the 3\% systematic error floor and $\ell_{\rm max}=30000$. \textit{Right}: SNR with the Gaussian covariance of Eq.~\eqref{eq:covmat} (grey) and with the 3\% error floor for $\ell_{\rm max}=8500$ (purple) and $\ell_{\rm max}=30000$ (green).}
            \label{fig:fisher_hd_A}
        \end{figure*}

        Figure~\ref{fig:fisher_hd_A} shows the forecast for case A. With the floor, the correlation matrix has the same block structure as for the \lact\,configuration. Within the foreground block, the amplitude and spectral index of radio sources remain strongly correlated ($\rho\simeq0.95$, compared with 0.99 for the \lact\,case). The pairs of parameters describing the same component of the CO occupation, i.e. $\alpha_{\rm CO}-M_{\rm sat}^{\rm CO}$ and $\sigma_{\rm CO}-\log M_{\rm min}^{\rm CO}$, are degenerate by construction, with $\rho\simeq-0.76$ and $\rho\simeq-0.87$ for the two pairs, respectively.

       The correlations between cosmological and foreground parameters are stronger than for the \lact\,configuration. The tSZ parameter $b_{HM}$ is correlated with $\mathcal{A}_s$, $\tau$, and the slope of the satellite occupation of CIB emitters, $\alpha_{\rm CIB}$. At the multipoles probed by CMB-HD, the total power is dominated by the CIB and, at lower frequencies, by the tSZ and radio sources. In our model, both CIB and tSZ depend on cosmology through the halo mass function, the halo bias, and the matter power spectrum. Changes in the amplitude and the shape of the matter power spectrum can therefore be partially compensated by the tSZ amplitude or by the occupation of CIB emitters.

        The right panel of Fig.~\ref{fig:fisher_hd_A} compares the SNR obtained with the Gaussian covariance (grey) and with the floor for the two different $\ell$ cuts. With Gaussian errors, all the parameters, including the four CO occupation parameters, would be constrained with SNR$>1$. The floor reduces the SNR. With $\ell_{\rm max}=30000$, the cosmological parameters are constrained with $\rm{SNR}\simeq10-40$, except for $\tau$ ($\rm{SNR}\simeq1$); the radio and SZ parameters are detected with high significance too; among the occupation parameters, only the minimum halo masses of CO and CIB emitters remain constrained. Restricting the analysis to $\ell_{\rm max}=8500$ leaves the SNR of the cosmological parameters almost unchanged but reduces that of the foreground parameters.

        The same results hold without CO satellites (case B, right panel of Fig.~\ref{fig:app_corr_nosat}).

    \subsection{Impact of neglecting CO}\label{subsec:noco}
        We now quantify what happens if CO is present in the data but not included in the model. We assume that the true sky contains CO, so that the covariance of Eq.~\eqref{eq:covmat} is computed with the total spectra including \coco\, and \cocib, and that the data are fitted with a model in which both the CO components and the CO parameters are removed. We denote by $\mathbf{F}_{\rm no CO}$ the Fisher matrix of this model, computed with Eq.~\eqref{eq:fisher} and the same covariance. To first order in the neglected signal, the parameters of the no-CO model are shifted by \citep{Amara:2008}
        \begin{equation}\label{eq:fisher_bias}
            \delta\theta_i=\sum_j\left(\mathbf{F}^{-1}_{\rm no\,CO}\right)_{ij}\,\sum_{bb^\prime}\dfrac{\partial C_b^{\rm no\,CO}}{\partial\theta_j}\,\Sigma^{-1}_{bb^\prime}\,\Delta C^{\rm CO}_{b^\prime} ,
        \end{equation}
        where $b,b^\prime$ run over the bandpowers of all the frequency combinations and $\Delta C_b^{\rm CO}$ is the CO contribution (\coco\,plus \cocib) to bandpower $b$. We report the bias in units of the uncertainty of the no-CO model, $\delta\theta_i/\sigma_{i,\rm noCO}$.

        \subsubsection{\lact\,configuration} 
\begin{figure}[h!]
            \centering
            \includegraphics[width=\linewidth]{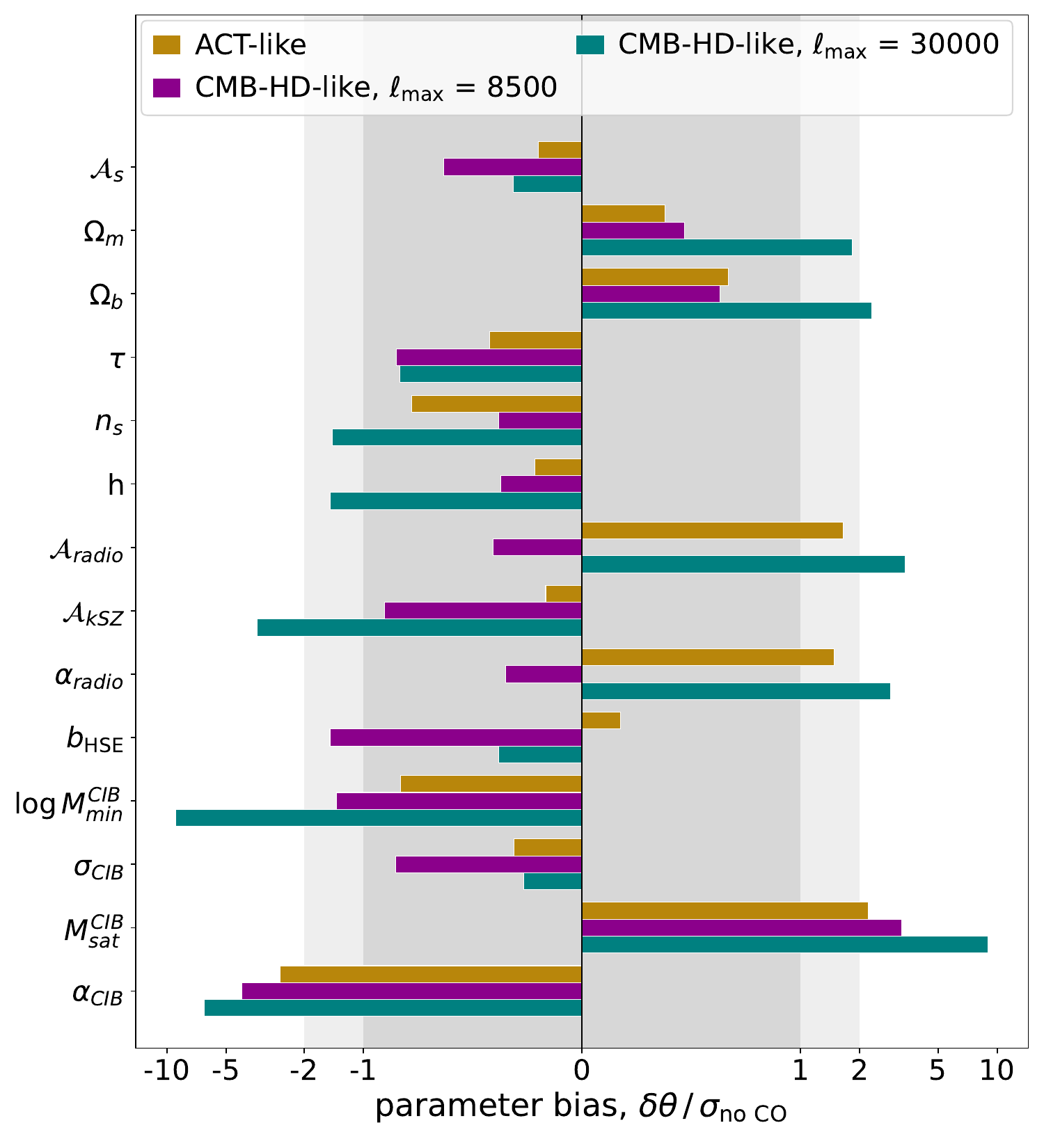}
            \caption{Bias induced on the parameters of the \lact\,(yellow) and of the \lcmnhd\,for $\ell_{\rm max}=8500$ and $\ell_{\rm max}=30000$ (purple and green, respectively), when CO is present in the data but neglected in the model, in units of the uncertaintiy of the no-CO model, $\delta\theta_i/\sigma_{i,\rm{noCO}}$. The dark and light grey bands mark the $1\sigma$ and $2\sigma$ intervals, respectively.}
            \label{fig:bias_act_hd}
        \end{figure}

The yellow bars of Fig.~\ref{fig:bias_act_hd} show the bias for case A. The cosmological parameters are essentially unaffected, with shifts below $0.8\sigma$ (see the left panel of Fig.~\ref{fig:act_triangles} for the Fisher constraints on the cosmological parameters). The foreground parameters, instead, absorb most of the CO power: the amplitude and the spectral index of radio sources are shifted by $\sim2\sigma$, and the tSZ and kSZ parameters by less than $\sim1\sigma$. In particular, a large part of the CO power is absorbed by the satellite occupation of the CIB emitters. Neglecting CO also tightens the constraints on the foreground parameters (see the right panel of Fig.~\ref{fig:act_triangles}).

        \subsubsection{Neglecting CO: \lcmnhd} For the \lcmnhd\,configuration, we compute the bias with the modification to the covariance of Sect.~\ref{subsub:cmbhdL}, i.e. with the 3\% floor, for $\ell_{\rm max}=8500$ and 30000 (purple and green bars of Fig.~\ref{fig:bias_act_hd}). In both cases, CO is detected with high significance. For $\ell_{\rm max}=8500$, all the cosmological parameters are shifted by less than $1\sigma$, while the satellite occupation of CIB emitters is biased by $\simeq3-4\sigma$, the tSZ parameter and the minimum halo mass of CIB emitters by $\simeq1\sigma$, and the kSZ and radio parameters by less than $1\sigma$. When all multipoles up to $\ell_{\rm max}=30000$ are used, the cosmological parameters are still shifted by $\leq1.5\sigma$, apart from $\Omega_m$ and $\Omega_b$ ($\simeq2\sigma$), while the bias increases for the foreground parameters that dominate the smallest scales. These shifts depend on the error budget assumed at the smallest scales: with a floor of 10\% the cosmological parameters are shifted by less than $0.8\sigma$ also for $\ell_{\rm max}=30000$ (Appendix~\ref{app:fisher})

        \begin{figure*}
            \centering
            \includegraphics[width=0.7\linewidth]{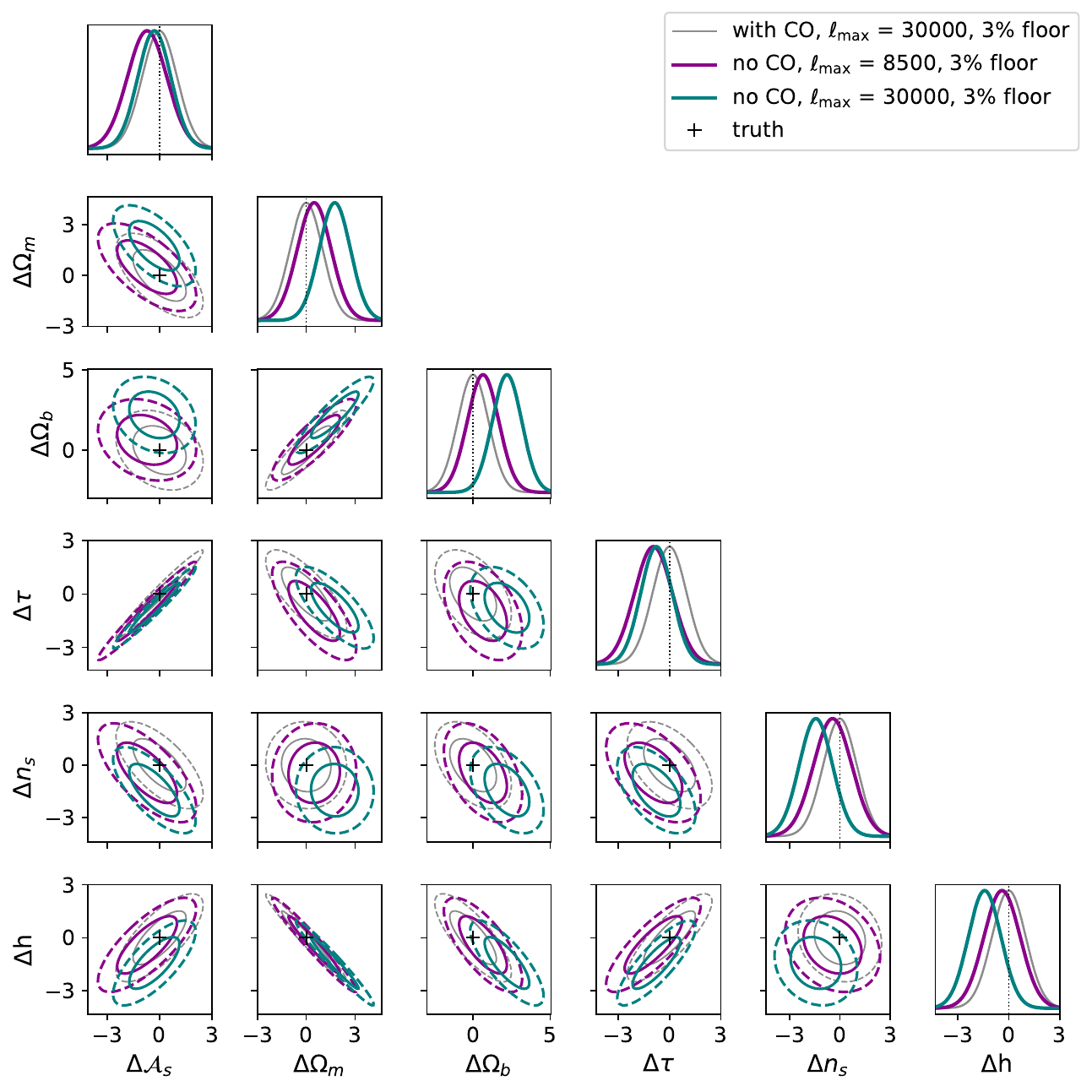}
            \caption{Forecast constraints on the cosmological parameters for the \lcmnhd\,configuration (case A) with the 3\% error floor, with CO in the model and $\ell_{\rm max}=30000$ (grey), and with CO neglected for $\ell_{\rm max}=8500$ and 30000 (purple and green, respectively). The axes show the shift from the fiducial value in units of the uncertainty with CO in the model; solid and dashed contours mark the 68\% and 95\% confidence regions, respectively.}
            \label{fig:hd_lmax_triangle}
        \end{figure*}
        Overall, for the error budgets considered here, neglecting CO in a \lcmnhd\,analysis biases the cosmological parameters by an amount that is small, but not negligible (see Fig.~\ref{fig:hd_lmax_triangle}): with the 3\% floor, the cosmological parameters are shifted by less than $1\sigma$ with the multipole range of the \lact\,configuration, and by up to $\simeq2\sigma$ when the smallest scales are included. With the 1\% floor, the shifts grow to $\simeq2\sigma$ even for $\ell_{\rm max}=8500$ (Appendix~\ref{app:fisher}, Fig.~\ref{fig:hd_bias_1pc}). Since CO remains detectable in all the configurations we considered, the most robust strategy is to include CO in the foreground model of future high-resolution CMB experiments.

\section{Discussion}\label{sec:discussion}
    In this section, we place the predictions of our model in the context of the small-scale CMB sky and of previous works, discuss what the Fisher forecasts imply for current and future CMB analyses, and summarise the limitations of the model. We focus on the assumptions specific to CO emission (SFR, SLED, satellite emission). The ingredients of the underlying halo model, i.e. the halo mass function, halo bias, density profile, and concentration-mass relation (Sect.~\ref{subsec:hm}), are fixed to standard choices and are not varied in this work.

    \subsection{CO in the arcminute-scale CMB sky}\label{subsec:discussion_cmbsky}
\begin{figure*}
            \centering
            \includegraphics[width=\linewidth]{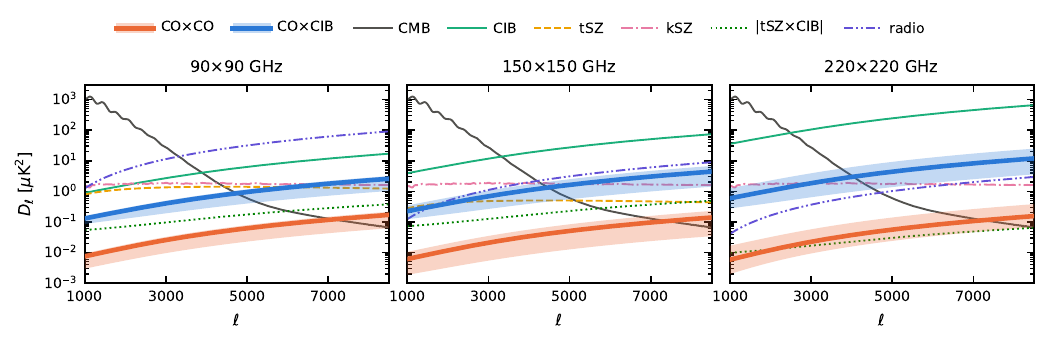}
            \caption{Contributions to the temperature power spectrum in the \lact\,auto-frequency channels. Thin lines: primary CMB, CIB, tSZ, kSZ, tSZ$\times$CIB, and radio sources. Thick lines: \coco\,(red) and \cocib\,(blue) for the baseline model; the shaded regions show the range spanned by the SLED and SFR prescriptions.}
            \label{fig:sky_components}
        \end{figure*}

        Figure~\ref{fig:sky_components} compares the \coco\,and \cocib\,power spectra predicted by our baseline model with the other components of the small-scale CMB sky: the primary CMB, CIB, tSZ, kSZ, tSZ$\times$CIB, and radio sources, computed with the foreground model of Appendix~\ref{app:fg_mod}, for the \lact\,auto-frequency channels. The shaded regions show the range spanned by the SLED and SFR prescriptions described in Sect.~\ref{subsec:co_lum}.

        In agreement with previous works \citep{Maniyar:2023, Kokron:2024, Mehta:2025}, the \cocib\, cross-spectrum is larger than the CO auto-spectrum in all channels. At $\ell=3000$, we find $D_{\ell=3000}^{CO\times CO}\simeq 2.6\times10^{-2},\,2.1\times10^{-2},\,2.3\times10^{-2}$ \muk and $D_{\ell=3000}^{CO\times CIB}\simeq 0.40,\,0.69,\,1.85$ \muk at 90, 150, and 220\ghz, respectively. The CO auto-spectrum is the smallest component in all channels, while the \cocib\,power grows with frequency. The latter is larger than the tSZ$\times$CIB power in all channels and comparable to the kSZ power at 220\ghz. Within the range allowed by the model assumptions, the \cocib\,power at 220\ghz\,can even exceed the kSZ one.

        Table~\ref{tab:co_auto_comparison} compares our predictions for both the \coco\, and the \cocib\, spectra at $\ell=3000$ with literature values. Our predictions are lower than those of \cite{Maniyar:2023} and \cite{Kokron:2024}, and lie in the lower part of the range spanned by the models of \cite{Mehta:2025}. We attribute this mainly to the different modelling of the CO luminosity. Our model describes the CO emission of main-sequence star-forming galaxies through the halo-averaged SFR of \um. It does not include the population of starburst galaxies that dominates the bright end of the observed CO luminosity functions; this population is instead captured by the empirical $L_{\rm IR}-L_{\rm CO}$ relations calibrated on observed galaxies and adopted by~\cite{Maniyar:2023},~\cite{Kokron:2024}, and~\cite{Mehta:2025}.
        \begin{table*}
            \centering
            \caption{Amplitudes $D_\ell$ of the auto-frequency \coco\ and \cocib\ spectra at $\ell=3000$ for common CMB observing frequencies, compared with previous works.}
            \label{tab:co_auto_comparison}
            \label{tab:cocib_auto_comparison}
            \begin{tabular}{c||c|c|c||c}
                \hline\hline
                Frequency & M23 & K24 & M25 & This work \\
                (GHz) & ($\mu$K$^2$) & ($\mu$K$^2$) & ($\mu$K$^2$) & ($\mu$K$^2$) \\
                \hline\hline
                \multicolumn{5}{c}{\textit{\coco}}\\
                \hline
                90  & --   & 0.05 -- 0.28 & 0.0054 -- 0.083 & 0.010 -- 0.032 \textbf{(0.026)} \\
                150 & 0.08 & 0.02 -- 0.2  & 0.0037 -- 0.061 & 0.006 -- 0.035 \textbf{(0.021)} \\
                220 & 0.5  & 0.08 -- 0.36 & 0.0049 -- 0.105 & 0.009 -- 0.058 \textbf{(0.023)} \\
                \hline
                \multicolumn{5}{c}{\textit{\cocib}}\\
                \hline
                90  & --   & 0.4 -- 1.1  & 0.34 -- 1.08 & 0.19 -- 0.49 \textbf{(0.40)} \\
                150 & 0.89 & 0.8 -- 2.3  & 0.55 -- 1.76 & 0.27 -- 1.06 \textbf{(0.69)} \\
                220 & 3.20 & 3.1 -- 7.2  & 1.07 -- 5.56 & 0.64 -- 3.9 \textbf{(1.85)} \\
                \hline
            \end{tabular}
            \tablefoot{M23: \citet{Maniyar:2023}; K24: \citet{Kokron:2024}; M25: \citet{Mehta:2025}. The values of M23, K24, and M25 are taken from Table~2 of \citet{Mehta:2025}. The ranges for this work cover the different SLED and SFR prescriptions, while the values in bold refer to the baseline.}
        \end{table*}

        Our Fisher results are in qualitative agreement with~\cite{Mehta:2025}, who found that neglecting \cocib\,in a CMB-S4-like analysis biases the foreground parameters; in our forecasts, too, the parameters most affected by neglecting CO are the foreground ones.

    \subsection{Implications for CMB analyses}
        The Fisher forecasts of Sect.~\ref{sec:results} show two different regimes.

        For an \lact\,experiment, CO cannot be constrained by the data: the CO occupation parameters have $\rm{SNR}\leq1$, and only the minimum halo mass of CIB emitters is constrained. However, CO cannot be ignored either. When it is neglected, the cosmological parameters are unaffected, but the radio parameters and the satellite occupation of CIB emitters absorb most of the CO power. Neglecting CO also yields tighter foreground constraints; including CO in the model broadens them without affecting the cosmological constraints. This is important whenever the foreground parameters are used to infer physical quantities, such as the duration and epoch of reionisation from the kSZ amplitude, the hydrostatic mass bias from the tSZ amplitude, or the galaxy-halo connection of dusty star-forming galaxies from the CIB occupation parameters: neglecting CO would yield constraints on these quantities that are both biased and too tight.

        For a \lcmnhd\,experiment, the Gaussian covariance underestimates the uncertainties and the results depend on the assumed error budget. With a 3\% systematic error floor, the minimum halo masses of both CO and CIB emitters are constrained, while the other occupation parameters are not, and the amplitude and spectral index of radio sources remain strongly correlated. The correlations between cosmological and foreground parameters are stronger than for the \lact\,case, because at high $\ell$ the foregrounds dominate the signal and, in our study, depend on cosmology through the halo model. CO is detected with high significance, and neglecting it shifts the cosmological parameters by less than $1\sigma$ for $\ell_{\rm max}=8500$ and by up to $2\sigma$ for the full CMB-HD multipole range. CO must therefore be part of the foreground model of \lcmnhd\,experiments.

        Our tests of Sect.~\ref{sec:mod_ass} also indicate which ingredients of the model matter most. The CO line intensity peaks at $z\simeq1.3$, with the mid-J transitions ($J=3-5$) dominating, so that each CMB channel receives contributions from a few transitions emitted at different redshifts. The main modelling uncertainties are the SFR prescription and the CO SLED, which change the amplitude of the CO power spectrum by a factor of $\sim 2-5$, depending on the channel, with little effect on its shape. Satellite galaxies, often neglected in LIM models, increase the \coco\,and the \cocib\,power by $\simeq40\%$ and $\simeq25\%$, respectively. By contrast, transitions above $J\simeq10$ change the \coco\,spectra by less than 2\% at frequencies up to 220\ghz, and haloes below $10^{10.5}M_\odot h^{-1}$ can be neglected. Truncating the model at $J_{\rm max}\simeq10$ and $M_{\rm low}=10^{10.5}M_\odot h^{-1}$ is therefore adequate for faster evaluations.

    \subsection{Limitations and future prospects}
        Our model has some limitations that we plan to address in future work. First, we adopted a single SLED for all galaxies, whereas the CO excitation is expected to vary with star formation and redshift. This can be addressed by replacing the constant coefficients $R_J$ of Eq.~\eqref{eq:lco} with a SLED that depends on halo mass and redshift, for instance by relating the CO excitation to the SFR surface density~\citep{Narayanan:2014ica, Daddi:2014mpa}. Second, the model describes the CO emission of main-sequence galaxies through the halo-averaged SFR, and does not include starburst galaxies. Starbursts can be included as a second emitting population, and this means splitting the luminosity of Eq.~\eqref{eq:lco} into a main-sequence and a starburst term, the latter with a more excited SLED; the M82-based normalisation of V10 is a natural starting point for this term. Third, our Fisher analysis assumes Gaussian covariances and white noise. For a \lcmnhd\,configuration, this approximation becomes important, and we have replaced the missing non-Gaussian and systematic contributions with an error floor; a realistic covariance, including the point-source trispectrum, is needed to assess the constraints on the CO parameters and the bias induced by CO more precisely. Finally, while we present results for \lact\,and \lcmnhd\,configurations, the framework applies to any frequency, angular scale, and experiment.

\section{Conclusions}\label{sec:conc}
    Rotational CO lines from star-forming galaxies are redshifted into the frequency bands of CMB experiments, where the superposition of many transitions emitted over wide redshift ranges produces a continuum-like foreground strongly correlated with the CIB. In this work, we developed an analytic, full-shape halo model for the \coco\,auto-spectrum and the \cocib\,cross-spectrum in CMB bands, in which the CO and infrared luminosities are derived from the same SFR, so that the correlation between CO and CIB is a natural prediction of the model.

    At $\ell=3000$, the \cocib\,contribution to the sky power is 0.4, 0.7, and 1.9\muk\,at 90, 150, and 220\ghz, comparable to the kSZ power at 220\ghz. The CO auto-spectrum is $\simeq0.02$\muk\,in all three channels. The SFR prescription and the CO SLED are the main sources of model uncertainty. For an \lact\,experiment, CO is not constrained by the data, but neglecting it biases both the radio parameters and the satellite occupation of CIB emitters. The cosmological parameters, instead, remain essentially unbiased (shifts below $0.8\sigma$): for the CO amplitude predicted by our baseline model, not modelling CO does not compromise the cosmological results of \lact\,analyses. For a \lcmnhd\,experiment, CO is detected, and neglecting it shifts the cosmological parameters by up to $\simeq2\sigma$. We did not perform a dedicated forecast for the Simons Observatory. Since its sensitivity lies between those of the two configurations considered here, we expect intermediate results, but a dedicated analysis is left for future work. CO should therefore be included in the foreground model of future high-resolution CMB analyses.

    The code implementing the model is publicly available\footnote{\href{https://github.com/giorgiazagatti/extragalacticCO_CIB.git}{\color{magenta}\faGithub\,\, giorgiazagatti/extragalacticCO\_CIB}}.

\begin{acknowledgements}
    We thank Alexander Van Engelen for useful discussions during the preparation of this work. GZ is funded by the European Union (ERC, RELiCS, project number 101116027) and by the \textit{Angelo Della Riccia foundation}. GZ acknowledges the financial support from the COSMOS network (www.cosmosnet.it) through the ASI (Italian Space Agency) Grants 2016-24-H.0 and 2016-24-H.1-2018. GZ also thanks the Institut d’Astrophysique Spatiale (IAS, Orsay, Université Paris-Saclay) for hosting GZ during this research period and for partial financial support from the French Agence Nationale de la Recherche (ANR) under grant ANR-22-CE31-0010 (project BATMAN). This work was also supported by CNES.
\end{acknowledgements}

\bibliographystyle{aa}
\bibliography{bibliography}

@article{Behroozi:2019,
        title = {UNIVERSEMACHINE: The correlation between galaxy growth and dark matter halo assembly from z = 0-10},
        author = {{Behroozi}, Peter and {Wechsler}, Risa H. and {Hearin}, Andrew P. and {Conroy}, Charlie},
      journal = {\mnras},
         year = 2019,
        month = sep,
       volume = {488},
       number = {3},
        pages = {3143-3194},
          doi = {10.1093/mnras/stz1182},
archivePrefix = {arXiv},
       eprint = {1806.07893},
 primaryClass = {astro-ph.GA},
       adsurl = {https://ui.adsabs.harvard.edu/abs/2019MNRAS.488.3143B}
}

@article{Tinker:2008ff,
    author = "Tinker, Jeremy L. and Kravtsov, Andrey V. and Klypin, Anatoly and Abazajian, Kevork and Warren, Michael S. and Yepes, Gustavo and Gottlober, Stefan and Holz, Daniel E.",
    title = "{Toward a halo mass function for precision cosmology: The Limits of universality}",
    eprint = "0803.2706",
    archivePrefix = "arXiv",
    primaryClass = "astro-ph",
    doi = "10.1086/591439",
    journal = "Astrophys. J.",
    volume = "688",
    pages = "709--728",
    year = "2008"
}

@article{Tinker:2010my,
    author = "Tinker, Jeremy L. and Robertson, Brant E. and Kravtsov, Andrey V. and Klypin, Anatoly and Warren, Michael S. and Yepes, Gustavo and Gottlober, Stefan",
    title = "{The Large Scale Bias of Dark Matter Halos: Numerical Calibration and Model Tests}",
    eprint = "1001.3162",
    archivePrefix = "arXiv",
    primaryClass = "astro-ph.CO",
    doi = "10.1088/0004-637X/724/2/878",
    journal = "Astrophys. J.",
    volume = "724",
    pages = "878--886",
    year = "2010"
}

@article{Tinker:2009jp,
    author = "Tinker, Jeremy L. and Wechsler, Risa H. and Zheng, Zheng",
    title = "{Interpreting the Clustering of Distant Red Galaxies}",
    eprint = "0902.1748",
    archivePrefix = "arXiv",
    primaryClass = "astro-ph.CO",
    reportNumber = "SLAC-PUB-13720",
    doi = "10.1088/0004-637X/709/1/67",
    journal = "Astrophys. J.",
    volume = "709",
    pages = "67--76",
    year = "2010"
}

@article{Navarro:1996gj,
    author = "Navarro, Julio F. and Frenk, Carlos S. and White, Simon D. M.",
    title = "{A Universal density profile from hierarchical clustering}",
    eprint = "astro-ph/9611107",
    archivePrefix = "arXiv",
    doi = "10.1086/304888",
    journal = "Astrophys. J.",
    volume = "490",
    pages = "493--508",
    year = "1997"
}

@article{Bullock:1999he,
    author = "Bullock, James S. and Kolatt, Tsafrir S. and Sigad, Yair and Somerville, Rachel S. and Kravtsov, Andrey V. and Klypin, Anatoly A. and Primack, Joel R. and Dekel, Avishai",
    title = "{Profiles of dark haloes. Evolution, scatter, and environment}",
    eprint = "astro-ph/9908159",
    archivePrefix = "arXiv",
    doi = "10.1046/j.1365-8711.2001.04068.x",
    journal = "Mon. Not. Roy. Astron. Soc.",
    volume = "321",
    pages = "559--575",
    year = "2001"
}

@article{Maniyar:2020tzw,
    author = "Maniyar, A. S. and B\'ethermin, M. and Lagache, G.",
    title = "{Simple halo model formalism for the cosmic infrared background and its correlation with the thermal Sunyaev-Zel\textquoteright{}dovich effect}",
    eprint = "2006.16329",
    archivePrefix = "arXiv",
    primaryClass = "astro-ph.CO",
    doi = "10.1051/0004-6361/202038790",
    journal = "Astron. Astrophys.",
    volume = "645",
    pages = "A40",
    year = "2021"
}

@article{Asgari:2023mej,
    author = "Asgari, Marika and Mead, Alexander J. and Heymans, Catherine",
    title = "{The halo model for cosmology: a pedagogical review}",
    eprint = "2303.08752",
    archivePrefix = "arXiv",
    primaryClass = "astro-ph.CO",
    doi = "10.21105/astro.2303.08752",
    month = "3",
    year = "2023"
}

@article{Tinker:2009mx,
    author = "Tinker, Jeremy L. and Wetzel, Andrew R.",
    title = "{What Does Clustering Tell Us About the Buildup of the Red Sequence?}",
    eprint = "0909.1325",
    archivePrefix = "arXiv",
    primaryClass = "astro-ph.CO",
    doi = "10.1088/0004-637X/719/1/88",
    journal = "Astrophys. J.",
    volume = "719",
    pages = "88--103",
    year = "2010"
}

@article{Lewis:1999bs,
      author         = "Lewis, Antony and Challinor, Anthony and Lasenby,
                        Anthony",
      title          = "{Efficient computation of CMB anisotropies in closed FRW
                        models}",
      journal        = "\apj",
      volume         = "538",
      year           = "2000",
      pages          = "473-476",
      doi            = "10.1086/309179",
      eprint         = "astro-ph/9911177",
      archivePrefix  = "arXiv",
      primaryClass   = "astro-ph",
      SLACcitation   = "%%CITATION = ASTRO-PH/9911177;%%"
}

@article{Howlett:2012mh,
      author         = "Howlett, Cullan and Lewis, Antony and Hall, Alex and
                        Challinor, Anthony",
      title          = "{CMB power spectrum parameter degeneracies in the era of
                        precision cosmology}",
      journal        = "\jcap",
      volume         = "1204",
      year           = "2012",
      pages          = "027",
      doi            = "10.1088/1475-7516/2012/04/027",
      eprint         = "1201.3654",
      archivePrefix  = "arXiv",
      primaryClass   = "astro-ph.CO",
      SLACcitation   = "%%CITATION = ARXIV:1201.3654;%%"
}

@article{Cooray:2002dia,
    author = "Cooray, Asantha and Sheth, Ravi K.",
    title = "{Halo Models of Large Scale Structure}",
    eprint = "astro-ph/0206508",
    archivePrefix = "arXiv",
    reportNumber = "FERMILAB-PUB-02-284-A",
    doi = "10.1016/S0370-1573(02)00276-4",
    journal = "Phys. Rept.",
    volume = "372",
    pages = "1--129",
    year = "2002"
}

@article{Dunkley:2013vu,
    author = "Dunkley, J. and others",
    title = "{The Atacama Cosmology Telescope: likelihood for small-scale CMB data}",
    eprint = "1301.0776",
    archivePrefix = "arXiv",
    primaryClass = "astro-ph.CO",
    doi = "10.1088/1475-7516/2013/07/025",
    journal = "JCAP",
    volume = "07",
    pages = "025",
    year = "2013"
}

@article{Zagatti:2024bhs,
    author = "Zagatti, Giorgia and Calabrese, Erminia and Chiocchetta, Caterina and Gerbino, Martina and Negrello, Mattia and Pagano, Luca",
    title = "{A halo model approach to describe clustering and emission of the two main star-forming galaxy populations for cosmic infrared background studies}",
    eprint = "2406.19069",
    archivePrefix = "arXiv",
    primaryClass = "astro-ph.CO",
    doi = "10.1051/0004-6361/202451424",
    journal = "Astron. Astrophys.",
    volume = "692",
    pages = "A190",
    year = "2024"
}

@article{George:2014oba,
    author = "George, E. M. and others",
    title = "{A measurement of secondary cosmic microwave background anisotropies from the 2500-square-degree SPT-SZ survey}",
    eprint = "1408.3161",
    archivePrefix = "arXiv",
    primaryClass = "astro-ph.CO",
    reportNumber = "FERMILAB-PUB-14-372-A",
    doi = "10.1088/0004-637X/799/2/177",
    journal = "Astrophys. J.",
    volume = "799",
    number = "2",
    pages = "177",
    year = "2015"
}

@article{Planck:2018vyg,
    author = "Aghanim, N. and others",
    collaboration = "Planck",
    title = "{Planck 2018 results. VI. Cosmological parameters}",
    eprint = "1807.06209",
    archivePrefix = "arXiv",
    primaryClass = "astro-ph.CO",
    doi = "10.1051/0004-6361/201833910",
    journal = "Astron. Astrophys.",
    volume = "641",
    pages = "A6",
    year = "2020",
    note = "[Erratum: Astron.Astrophys. 652, C4 (2021)]"
}

@article{SimonsObservatory:2018koc,
    author = "Ade, Peter and others",
    collaboration = "Simons Observatory",
    title = "{The Simons Observatory: Science goals and forecasts}",
    eprint = "1808.07445",
    archivePrefix = "arXiv",
    primaryClass = "astro-ph.CO",
    doi = "10.1088/1475-7516/2019/02/056",
    journal = "JCAP",
    volume = "02",
    pages = "056",
    year = "2019"
}

@article{Visbal2010,
       author = "{Visbal}, Eli and {Loeb}, Abraham",
        title = "{Measuring the 3D clustering of undetected galaxies through cross correlation of their cumulative flux fluctuations from multiple spectral lines}",
      journal = "\jcap",
         year = "2010",
        month = "nov",
       volume = "2010",
       number = "11",
          eid = "016",
        pages = "016",
          doi = "10.1088/1475-7516/2010/11/016",
archivePrefix = "arXiv",
       eprint = "1008.3178",
 primaryClass = "astro-ph.CO"
}

@article{Righi:2008,
    author = "Righi, M. and Hern{\'a}ndez-Monteagudo, C. and Sunyaev, R. A.",
    title = "{Carbon monoxide line emission as a CMB foreground: tomography of the star-forming universe with different spectral resolutions}",
    journal = "\aap",
    volume = "489",
    pages = "489--504",
    year = "2008"
}

@article{Maniyar:2023,
    author = "Maniyar, Abhishek S. and Gkogkou, Athanasia and Coulton, William R. and Li, Zhiqi and Lagache, Guilaine and Pullen, Anthony R.",
    title = "{Extragalactic CO emission lines in the CMB experiments: A forgotten signal and a foreground}",
    journal = "\prd",
    volume = "107",
    number = "12",
    pages = "123504",
    year = "2023",
    doi = "10.1103/PhysRevD.107.123504",
    eprint = "2301.10764",
    archivePrefix = "arXiv"
}

@article{Kokron:2024,
    author = "Kokron, Nickolas and Bernal, Jos{\'e} Luis and Dunkley, Jo",
    title = "{Contributions of extragalactic CO emission lines to ground-based CMB observations}",
    journal = "\prd",
    volume = "110",
    pages = "103535",
    year = "2024",
    doi = "10.1103/PhysRevD.110.103535",
    eprint = "2405.20369",
    archivePrefix = "arXiv"
}

@article{Mehta:2025,
    author = "Mehta, Yogesh and Roy, Anirban and Foreman, Simon and van Engelen, Alexander and Battaglia, Nicholas",
    title = "{The Modeling Landscape of Extragalactic CO in CMB Surveys}",
    journal = "arXiv e-prints",
    year = "2025",
    eprint = "2506.16028",
    archivePrefix = "arXiv",
    primaryClass = "astro-ph.CO"
}

@article{Roy:2023limpy,
    author = "Roy, Anirban and Valent{\'i}n-Mart{\'i}nez, Dariannette and Wang, Kailai and Battaglia, Nicholas and van Engelen, Alexander",
    title = "{LIMpy: A Semianalytic Approach to Simulating Multiline Intensity Maps at Millimeter Wavelengths}",
    journal = "\apj",
    volume = "957",
    pages = "87",
    year = "2023",
    doi = "10.3847/1538-4357/acf92f",
    eprint = "2304.06748",
    archivePrefix = "arXiv"
}

@article{Roy:2024CO32,
    author = "Roy, Anirban and Battaglia, Nicholas and Pullen, Anthony R.",
    title = "{A Measurement of CO(3-2) Line Emission from eBOSS Galaxies at $z\sim0.5$ using Planck Data}",
    journal = "arXiv e-prints",
    year = "2024",
    eprint = "2406.07861",
    archivePrefix = "arXiv",
    primaryClass = "astro-ph.GA"
}

@article{Silva:2015,
    author = "Silva, Marta and Santos, Mario G. and Cooray, Asantha and Gong, Yan",
    title = "{Prospects for Detecting C II Emission during the Epoch of Reionization}",
    journal = "\apj",
    volume = "806",
    pages = "209",
    year = "2015",
    doi = "10.1088/0004-637X/806/2/209",
    eprint = "1410.4808",
    archivePrefix = "arXiv"
}

@article{Fonseca:2017,
    author = "Fonseca, Jos{\'e} and Silva, Marta B. and Santos, M{\'a}rio G. and Cooray, Asantha",
    title = "{Cosmology with intensity mapping techniques using atomic and molecular lines}",
    journal = "\mnras",
    volume = "464",
    pages = "1948--1965",
    year = "2017",
    doi = "10.1093/mnras/stw2470",
    eprint = "1607.05288",
    archivePrefix = "arXiv"
}

@article{Kennicutt:1998,
    author = "Kennicutt, Jr., Robert C.",
    title = "{Star Formation in Galaxies Along the Hubble Sequence}",
    journal = "\araa",
    volume = "36",
    pages = "189--232",
    year = "1998",
    doi = "10.1146/annurev.astro.36.1.189",
    eprint = "astro-ph/9807187",
    archivePrefix = "arXiv"
}

@article{Limber:1953,
    author = "Limber, D. Nelson",
    title = "{The Analysis of Counts of the Extragalactic Nebulae in Terms of a Fluctuating Density Field.}",
    journal = "\apj",
    volume = "117",
    pages = "134",
    year = "1953",
    doi = "10.1086/145672"
}

@article{LoVerde:2008,
    author = "LoVerde, Marilena and Afshordi, Niayesh",
    title = "{Extended Limber approximation}",
    journal = "\prd",
    volume = "78",
    pages = "123506",
    year = "2008",
    doi = "10.1103/PhysRevD.78.123506",
    eprint = "0809.5112",
    archivePrefix = "arXiv"
}

@article{Carilli:2013,
    author = "Carilli, C. L. and Walter, F.",
    title = "{Cool Gas in High-Redshift Galaxies}",
    journal = "\araa",
    volume = "51",
    pages = "105--161",
    year = "2013",
    doi = "10.1146/annurev-astro-082812-140953",
    eprint = "1301.0371",
    archivePrefix = "arXiv"
}

@article{Bolatto:2013,
    author = "Bolatto, Alberto D. and Wolfire, Mark and Leroy, Adam K.",
    title = "{The CO-to-H$_2$ Conversion Factor}",
    journal = "\araa",
    volume = "51",
    pages = "207--268",
    year = "2013",
    doi = "10.1146/annurev-astro-082812-140944",
    eprint = "1301.3498",
    archivePrefix = "arXiv"
}

@article{Kovetz:2017,
    author = "Kovetz, Ely D. and others",
    title = "{Line-Intensity Mapping: 2017 Status Report}",
    journal = "arXiv e-prints",
    year = "2017",
    eprint = "1709.09066",
    archivePrefix = "arXiv",
    primaryClass = "astro-ph.CO"
}

@article{Bernal:2022,
    author = "Bernal, Jos{\'e} Luis and Kovetz, Ely D.",
    title = "{Line-intensity mapping: theory review with a focus on star-formation lines}",
    journal = "\aapr",
    volume = "30",
    pages = "5",
    year = "2022",
    doi = "10.1007/s00159-022-00143-0",
    eprint = "2206.15377",
    archivePrefix = "arXiv"
}

@article{Pullen:2013,
    author = "Pullen, Anthony R. and Chang, Tzu-Ching and Dor{\'e}, Olivier and Lidz, Adam",
    title = "{Cross-correlations as a Cosmological Carbon Monoxide Detector}",
    journal = "\apj",
    volume = "768",
    pages = "15",
    year = "2013",
    doi = "10.1088/0004-637X/768/1/15",
    eprint = "1211.1397",
    archivePrefix = "arXiv"
}

@article{Breysse:2014,
    author = "Breysse, Patrick C. and Kovetz, Ely D. and Kamionkowski, Marc",
    title = "{Carbon monoxide intensity mapping at moderate redshifts}",
    journal = "\mnras",
    volume = "443",
    pages = "3506--3512",
    year = "2014"
}

@article{Keating:2016,
    author = "Keating, Garrett K. and Marrone, Daniel P. and Bower, Geoffrey C. and Leitch, Erik and Carlstrom, John E. and DeBoer, David R.",
    title = "{COPSS II: The Molecular Gas Content of Ten Million Cubic Megaparsecs at Redshift z $\sim$ 3}",
    journal = "\apj",
    volume = "830",
    pages = "34",
    year = "2016",
    doi = "10.3847/0004-637X/830/1/34",
    eprint = "1605.03971",
    archivePrefix = "arXiv"
}

@article{Reichardt:2021,
    author = "Reichardt, C. L. and others",
    title = "{An Improved Measurement of the Secondary Cosmic Microwave Background Anisotropies from the SPT-SZ + SPTpol Surveys}",
    journal = "\apj",
    volume = "908",
    pages = "199",
    year = "2021",
    eprint = "2002.06197",
    archivePrefix = "arXiv"
}

@article{Louis:2025,
    author = "Louis, Thibaut and others",
    title = "{The Atacama Cosmology Telescope: DR6 Power Spectra, Likelihoods and $\Lambda$CDM Parameters}",
    journal = "arXiv e-prints",
    year = "2025",
    eprint = "2503.14452",
    archivePrefix = "arXiv",
    primaryClass = "astro-ph.CO"
}

@article{CMBHD:2022,
    author = "{CMB-HD Collaboration} and Aiola, Simone and others",
    title = "{Snowmass2021 CMB-HD White Paper}",
    journal = "arXiv e-prints",
    year = "2022",
    eprint = "2203.05728",
    archivePrefix = "arXiv",
    primaryClass = "astro-ph.CO"
}

@article{Sehgal:2019,
    author = "Sehgal, Neelima and others",
    title = "{CMB-HD: An Ultra-Deep, High-Resolution Millimeter-Wave Survey Over Half the Sky}",
    journal = "Bulletin of the American Astronomical Society",
    volume = "51",
    number = "7",
    pages = "6",
    year = "2019",
    eprint = "1906.10134",
    archivePrefix = "arXiv"
}

@article{Tristram:2025,
    author = "Tristram, M. and Douspis, M. and Gorce, A. and Henrot-Versill{\'e}, S. and Hergt, L. T. and Ilic, S. and McBride, L. and Mu{\~n}oz-Echeverr{\'i}a, M. and Pointecouteau, E. and Salvati, L.",
    title = "{Combining cosmic microwave background datasets with consistent foreground modelling}",
    journal = "arXiv e-prints",
    year = "2025",
    eprint = "2511.04733",
    archivePrefix = "arXiv",
    primaryClass = "astro-ph.CO"
}

@article{Zheng:2005,
    author = "Zheng, Zheng and others",
    title = "{Theoretical Models of the Halo Occupation Distribution: Separating Central and Satellite Galaxies}",
    journal = "\apj",
    volume = "633",
    pages = "791--809",
    year = "2005",
    doi = "10.1086/466510",
    eprint = "astro-ph/0408564",
    archivePrefix = "arXiv"
}

@article{Tegmark:1997,
    author = "Tegmark, Max and Taylor, Andy N. and Heavens, Alan F.",
    title = "{Karhunen-Lo{\`e}ve Eigenvalue Problems in Cosmology: How Should We Tackle Large Data Sets?}",
    journal = "\apj",
    volume = "480",
    pages = "22--35",
    year = "1997",
    doi = "10.1086/303939",
    eprint = "astro-ph/9603021",
    archivePrefix = "arXiv"
}

@article{Madau:2014,
    author = "Madau, Piero and Dickinson, Mark",
    title = "{Cosmic Star-Formation History}",
    journal = "\araa",
    volume = "52",
    pages = "415--486",
    year = "2014",
    doi = "10.1146/annurev-astro-081811-125615",
    eprint = "1403.0007",
    archivePrefix = "arXiv"
}

@article{Amara:2008,
    author = "Amara, Adam and R{\'e}fr{\'e}gier, Alexandre",
    title = "{Systematic bias in cosmic shear: extending the Fisher matrix}",
    journal = "\mnras",
    volume = "391",
    pages = "228--236",
    year = "2008",
    eprint = "0710.5171",
    archivePrefix = "arXiv"
}

@article{Tacconi:2020,
    author = "Tacconi, Linda J. and Genzel, Reinhard and Sternberg, Amiel",
    title = "{The Evolution of the Star-Forming Interstellar Medium Across Cosmic Time}",
    journal = "\araa",
    volume = "58",
    pages = "157--203",
    year = "2020",
    eprint = "2003.06245",
    archivePrefix = "arXiv"
}

@article{Solomon:2005,
    author = "Solomon, P. M. and Vanden Bout, P. A.",
    title = "{Molecular Gas at High Redshift}",
    journal = "\araa",
    volume = "43",
    pages = "677--725",
    year = "2005",
    eprint = "astro-ph/0508481",
    archivePrefix = "arXiv"
}

@article{Tacconi:2018,
    author = "Tacconi, L. J. and Genzel, R. and Saintonge, A. and others",
    title = "{PHIBSS: Unified Scaling Relations of Gas Depletion Time and Molecular Gas Fractions}",
    journal = "\apj",
    volume = "853",
    pages = "179",
    year = "2018",
    eprint = "1702.01140",
    archivePrefix = "arXiv"
}

@article{Decarli:2020,
    author = "Decarli, Roberto and others",
    title = "{The ALMA Spectroscopic Survey in the HUDF: Multiband Constraints on Line-luminosity Functions and the Cosmic Density of Molecular Gas}",
    journal = "\apj",
    volume = "902",
    pages = "110",
    year = "2020",
    eprint = "2009.10744",
    archivePrefix = "arXiv"
}

@article{Boogaard:2020,
    author = "Boogaard, Leindert A. and others",
    title = "{The ALMA Spectroscopic Survey in the HUDF: CO Excitation and Atomic Carbon in Star-forming Galaxies at z = 1-3}",
    journal = "\apj",
    volume = "902",
    pages = "109",
    year = "2020",
    eprint = "2009.04348",
    archivePrefix = "arXiv"
}

@article{Planck:2011ivn,
    author = "Ade, P. A. R. and others",
    collaboration = "Planck",
    title = "{Planck Early Results XVIII: The power spectrum of cosmic infrared background anisotropies}",
    eprint = "1101.2028",
    archivePrefix = "arXiv",
    primaryClass = "astro-ph.CO",
    doi = "10.1051/0004-6361/201116461",
    journal = "Astron. Astrophys.",
    volume = "536",
    pages = "A18",
    year = "2011"
}

@article{Klypin:2014kpa,
    author = "Klypin, Anatoly and Yepes, Gustavo and Gottlober, Stefan and Prada, Francisco and Hess, Steffen",
    title = "{MultiDark simulations: the story of dark matter halo concentrations and density profiles}",
    eprint = "1411.4001",
    archivePrefix = "arXiv",
    primaryClass = "astro-ph.CO",
    doi = "10.1093/mnras/stw248",
    journal = "Mon. Not. Roy. Astron. Soc.",
    volume = "457",
    number = "4",
    pages = "4340--4359",
    year = "2016"
}

@article{Douspis:2021ing,
    author = "Douspis, Marian and Salvati, Laura and Gorce, Ad{\'e}lie and Aghanim, Nabila",
    title = "{Retrieving cosmological information from small-scale CMB foregrounds. I. The thermal Sunyaev Zel'dovich effect}",
    journal = "\aap",
    volume = "659",
    pages = "A99",
    year = "2022",
    doi = "10.1051/0004-6361/202142004",
    eprint = "2109.03272",
    archivePrefix = "arXiv"
}

@article{Gorce:2022cvb,
    author = "Gorce, Ad{\'e}lie and Douspis, Marian and Salvati, Laura",
    title = "{Retrieving cosmological information from small-scale CMB foregrounds. II. The kinetic Sunyaev Zel'dovich effect}",
    journal = "\aap",
    volume = "662",
    pages = "A122",
    year = "2022",
    doi = "10.1051/0004-6361/202243351",
    eprint = "2202.08698",
    archivePrefix = "arXiv"
}

@article{Narayanan:2014ica,
    author = "Narayanan, Desika and Krumholz, Mark R.",
    title = "{A theory for the excitation of CO in star-forming galaxies}",
    journal = "\mnras",
    volume = "442",
    pages = "1411--1428",
    year = "2014",
    doi = "10.1093/mnras/stu834",
    eprint = "1401.2998",
    archivePrefix = "arXiv"
}

@article{Daddi:2014mpa,
    author = "Daddi, E. and Dannerbauer, H. and Liu, D. and Aravena, M. and Bournaud, F. and Walter, F. and Riechers, D. and Magdis, G. and Sargent, M. and B{\'e}thermin, M. and Carilli, C. and Cibinel, A. and Dickinson, M. and Elbaz, D. and Gao, Y. and Gobat, R. and Hodge, J. and Krips, M.",
    title = "{CO excitation of normal star-forming galaxies out to z = 1.5 as regulated by the properties of their interstellar medium}",
    journal = "\aap",
    volume = "577",
    pages = "A46",
    year = "2015",
    doi = "10.1051/0004-6361/201425043",
    eprint = "1409.8158",
    archivePrefix = "arXiv"
}

@article{Bernal:2019jdo,
    author = "Bernal, Jos{\'e} Luis and Breysse, Patrick C. and Gil-Mar{\'\i}n, H{\'e}ctor and Kovetz, Ely D.",
    title = "{User{\textquoteright}s guide to extracting cosmological information from line-intensity maps}",
    eprint = "1907.10067",
    archivePrefix = "arXiv",
    primaryClass = "astro-ph.CO",
    doi = "10.1103/PhysRevD.100.123522",
    journal = "Phys. Rev. D",
    volume = "100",
    number = "12",
    pages = "123522",
    year = "2019"
}

@article{MoradinezhadDizgah:2021dei,
    author = "Moradinezhad Dizgah, Azadeh and Nikakhtar, Farnik and Keating, Garrett K. and Castorina, Emanuele",
    title = "{Precision tests of CO and  CII  power spectra models against simulated intensity maps}",
    eprint = "2111.03717",
    archivePrefix = "arXiv",
    primaryClass = "astro-ph.CO",
    doi = "10.1088/1475-7516/2022/02/026",
    journal = "JCAP",
    volume = "02",
    number = "02",
    pages = "026",
    year = "2022"
}

@article{ACT:2023dou,
    author = "Qu, Frank J. and others",
    collaboration = "ACT",
    title = "{The Atacama Cosmology Telescope: A Measurement of the DR6 CMB Lensing Power Spectrum and Its Implications for Structure Growth}",
    eprint = "2304.05202",
    archivePrefix = "arXiv",
    primaryClass = "astro-ph.CO",
    reportNumber = "FERMILAB-PUB-23-237-PPD, FERMILAB-PUB-23-237-PPD",
    doi = "10.3847/1538-4357/acfe06",
    journal = "Astrophys. J.",
    volume = "962",
    number = "2",
    pages = "112",
    year = "2024"
}

@article{Battaglia:2010tm,
    author = "Battaglia, N. and Bond, J. R. and Pfrommer, C. and Sievers, J. L. and Sijacki, D.",
    title = "{Simulations of the Sunyaev-Zel'dovich Power Spectrum with Active Galactic Nucleus Feedback}",
    eprint = "1003.4256",
    archivePrefix = "arXiv",
    primaryClass = "astro-ph.CO",
    doi = "10.1088/0004-637X/725/1/91",
    journal = "Astrophys. J.",
    volume = "725",
    number = "1",
    pages = "91",
    year = "2010"
}

\appendix
\section{Halo model calculations}\label{app:hm_calc}
    In this appendix, we derive the halo model expressions for the \coco\,and the \cocib\,power spectra of Sects.~\ref{subsec:coco} and~\ref{subsec:cocib}.

    \subsection{Luminosity-weighted fields}
        For a halo of mass $m$ at position $\mathbf{x}_h$, hosting $N_c\in\{0,1\}$ central galaxies at its centre and $N_s$ satellite galaxies at positions $\mathbf{x}_h+\mathbf{r}_i$, the luminosity density of transition $J$ contributed by this halo is
        \begin{equation}
            \rho_J(\mathbf{x}) = N_c\,\ell_c^J\,\delta_{\rm D}(\mathbf{x}-\mathbf{x}_h) + \sum_{i=1}^{N_s}\ell_s^J\,\delta_{\rm D}(\mathbf{x}-\mathbf{x}_h-\mathbf{r}_i) ,
        \end{equation}
        where $\ell_{c/s}^J(m,z)=R_J\rm{SFR}_{c/s}(m,z)$ is the luminosity of a single central/satellite galaxy. In Fourier space,
        \begin{equation}
            \rho_J(\mathbf{k}) = e^{-i\mathbf{k}\cdot\mathbf{x}_h}\Bigl[N_c\,\ell_c^J + \ell_s^J\sum_{i=1}^{N_s}e^{-i\mathbf{k}\cdot\mathbf{r}_i}\Bigr] .
        \end{equation}
        We assume that satellite galaxies are distributed within the halo according to a normalised profile whose Fourier transform is $\tilde{u}_{\rm gal}(k,m)$, with no correlation between the positions of different satellites. Averaging over the satellite positions then gives $\langle e^{-i\mathbf{k}\cdot\mathbf{r}_i}\rangle=\tilde{u}_{\rm gal}(k,m)$ and, for $i\neq j$, $\langle e^{-i\mathbf{k}\cdot(\mathbf{r}_i-\mathbf{r}_j)}\rangle=\tilde{u}_{\rm gal}^2(k,m)$. The mean luminosity density is $\overline{j}_J=\int{\rm d}m\,({\rm d}n/{\rm d}m)\,[\langle N_c\rangle\ell_c^J+\langle N_s\rangle\ell_s^J]$ (Eq.~\eqref{eq:jbar}), and we work with the fractional fluctuation $\delta_J=\rho_J/\overline{j}_J-1$.

    \subsection{1h and shot noise terms}
        The 1h term is the average over haloes of the product of the Fourier amplitudes of the two fields within the same halo. For the  transitions $J$ and $J^\prime$,
        \begin{equation}
            \begin{aligned}
            \bigl\langle\rho_J\rho_{J^\prime}^*\bigr\rangle_{\rm halo} &= \langle N_c^2\rangle\,\ell_c^J\ell_c^{J^\prime} + \langle N_cN_s\rangle\bigl(\ell_c^J\ell_s^{J^\prime}+\ell_s^J\ell_c^{J^\prime}\bigr)\tilde{u}_{\rm gal}\\
            &+ \langle N_s\rangle\,\ell_s^J\ell_s^{J^\prime} + \langle N_s(N_s-1)\rangle\,\ell_s^J\ell_s^{J^\prime}\,\tilde{u}_{\rm gal}^2 .
            \end{aligned}
        \end{equation}
        The first and third terms correspond to the emission of the same galaxy in the two transitions, i.e. to pairs at zero separation. They are scale independent, and, since $N_c^2=N_c$, together they give the shot noise term of Eq.~\eqref{eq:Pshot_coco}. The remaining terms give the 1h term of Eq.~\eqref{eq:P1h_coco}, after integrating over the halo mass function and dividing by $\overline{j}_J\overline{j}_{J^\prime}$. In Eq.~\eqref{eq:P1h_coco}, we assume that the occupations of centrals and satellites are independent, $\langle N_cN_s\rangle=\langle N_c\rangle\langle N_s\rangle$, and that satellites follow Poisson statistics, for which $\langle N_s(N_s-1)\rangle=\langle N_s\rangle^2$.

    \subsection{2h term}
        On large scales, the halo centres trace the linear matter density field with bias $b(m)$. The 2h term is the linear matter power spectrum multiplied by the luminosity- and bias-weighted mean profiles of the two fields,
        \begin{equation}
            \begin{aligned}
                P^{2h}_{JJ^\prime}(k)=P^{\rm lin}_{mm}(k)\prod_{X=J,J^\prime}&\left[\dfrac{1}{\overline{j}_X}\int{\rm d}m\,\dfrac{{\rm d}n}{{\rm d}m}\,b(m)\right.\\
                &\times\left.\bigl(\langle N_c\rangle\ell_c^X+\langle N_s\rangle\ell_s^X\,\tilde{u}_{\rm gal}\bigr)\vphantom{\dfrac{1}{\overline{j}_X}}\right] ,
            \end{aligned}
        \end{equation}
        which is Eq.~\eqref{eq:P2h_coco}.

    \subsection{CO$\times$CIB}
        The \cocib\,spectrum follows from the same calculation by replacing one of the two CO fields with the infrared luminosity density of galaxies in the channel $\nu$, with single-galaxy luminosities $\ell_{c/s}^{\rm CIB}$. The occupations of CO and CIB emitters is described by two HODs, $N_{c/s}^{\rm CO}$ and $N_{c/s}^{\rm CIB}$, this gives the 1h and 2h terms of Eqs.~\eqref{eq:P1h_cocib} and~\eqref{eq:P2h_cocib}. The shot noise term arises from galaxies that emit in both CO and the infrared, whose mean number per halo is $N_{c/s}^{CO\cap CIB}=min\{N_{c/s}^{CO}, N_{c/s}^{CIB}\}$.

\section{Star formation rates of central and satellite galaxies}\label{app:um_sfr}
    In this appendix, we describe how we derive the SFR of central and satellite galaxies, ${\rm SFR}_c$ and ${\rm SFR}_s$, from \um\,\citep{Behroozi:2019}, what the satellite HOD describes once ${\rm SFR}_s$ is fixed, and how we include the scatter of the SFR at fixed halo mass. The procedure is the same for CO and CIB emitters, since the luminosities of both are derived from the same SFR.

    \subsection{Moments of the \um\,SFR distribution}
        We use the \um\,galaxy catalogues built on the Small MultiDark-Planck (SMDPL) simulation~\citep{Klypin:2014kpa}, in which each galaxy is either a central, hosted by a distinct halo, or a satellite, hosted by a subhalo and assigned to its top-level host halo. We bin the host haloes in mass and, in each bin, we compute the following averages over host haloes:
        \begin{equation}\label{eq:um_moments}
            \overline{\rm SFR}_c^{\rm UM}=\bigl\langle{\rm SFR}_c\bigr\rangle ,\quad S_1=\Bigl\langle\sum_{i}{\rm SFR}_i\Bigr\rangle ,\quad S_2=\Bigl\langle\sum_{i}{\rm SFR}_i^2\Bigr\rangle ,
        \end{equation}
        where ${\rm SFR}_c$ is the SFR of the central galaxy, and the sums run over all the satellites of the host halo. Here, $S_1$ is the mean total SFR in satellites, i.e. ${\rm SFR}_{s,\rm tot}^{\rm UM}$ of Sect.~\ref{subsec:coco}, while $S_2$ carries the information on how this SFR is distributed among the satellites.

    \subsection{Central and satellite SFR}
        Our baseline SFR, ${\rm SFR}(m,z)$, is the smooth B19 relation, which gives the total SFR of a halo. We split it into the central SFR, ${\rm SFR}_c=(1-f_{\rm sat})\,{\rm SFR}$, and the total SFR in satellites, $f_{\rm sat}\,{\rm SFR}$, with the \um\,satellite fraction of Sect.~\ref{subsec:coco}.

        The halo model also needs the SFR of a single satellite. For this, we define the following two quantities:
        \begin{equation}\label{eq:neff}
            {\rm SFR}_s=\dfrac{S_2}{S_1} ,\qquad N_s^{\rm eff}=\dfrac{S_1^2}{S_2} .
        \end{equation}
        Here, ${\rm SFR}_s$ is the SFR-weighted mean SFR of the satellites, and $N_s^{\rm eff}$ is the effective number of star-forming satellites. If all the satellites had the same SFR, $N_s^{\rm eff}$ would equal their number. In \um, however, a few bright satellites carry most of the satellite star formation

        To keep the normalisation of the B19 SFR, we take from \um\,only the shape of the satellite SFR distribution, $Q_s\equiv 1/N_s^{\rm eff}$, and assign to each satellite
        \begin{equation}\label{eq:sfr_sat}
            {\rm SFR}_s(m,z)=f_{\rm sat}(m,z)\,{\rm SFR}(m,z)\,Q_s(m,z) .
        \end{equation}
        At this point, ${\rm SFR}_s$ does not depend on any HOD parameter and is the same for CO and CIB emitters.

    \subsection{The satellite HOD}
        Given ${\rm SFR}_s$, the satellite contribution of a tracer $X$ (CO or CIB) is set by its satellite occupation $N_s^X(m)$ given by Eq.~\eqref{eq:N_sat}. We can write $N_s^X(m)$ in terms of $N_s^{\rm eff}$ as $N_s^X=B^X\,N_s^{\rm eff}$, where $B^X$ is the fraction of \um\,satellite star formation seen by tracer $X$. Equation~\eqref{eq:sfr_sat} gives
        \begin{equation}
            N_s^X\,{\rm SFR}_s=B^X\,f_{\rm sat}\,{\rm SFR} .
        \end{equation}
        For $B^X<1$, the tracer sees only part of the \um\,satellite star formation. Values $B^X>1$ would imply more satellite star formation than \um\,provides. We therefore require $B^{X}\leq1$, i.e. $N_s^X\leq N_s^{\rm eff}$, and the HOD thus counts effective, SFR-weighted emitters rather than galaxies.

\section{Emission kernels}\label{app:kernels}
    In this appendix, we derive the emission kernels of Eqs.~\eqref{eq:emiss_kernel} and \eqref{eq:cib_kernel}.

    \subsection{CO}
        The specific intensity at the observed frequency $\nu_{\rm obs}$ produced by the transition $J$ is, from Eq.~\eqref{eq:line_int},
        \begin{equation}
            I_{\nu}^J(\nu_{\rm obs}) = \dfrac{c}{4\pi}\dfrac{\overline{j}_J(z)}{\nu_J\,H(z)}, \qquad 1+z=\dfrac{\nu_J}{\nu_{\rm obs}} ,
        \end{equation}
        where $\overline{j}_J$ is the comoving luminosity density of Eq.~\eqref{eq:jbar}. A channel with spectral response $\mathcal{R}_\nu(\nu_{\rm obs})$ and width $\Delta\nu=\int{\rm d}\nu_{\rm obs}\,\mathcal{R}_\nu(\nu_{\rm obs})$ measures the band-averaged intensity
        \begin{equation}
            \overline{I}_{\nu}^J = \dfrac{1}{\Delta\nu}\int{\rm d}\nu_{\rm obs}\,\mathcal{R}_\nu(\nu_{\rm obs})\,I_{\nu}^J(\nu_{\rm obs}) .
        \end{equation}
        Changing variable to the emission redshift, ${\rm d}\nu_{\rm obs}=\nu_J\,{\rm d}z/(1+z)^2$, we obtain
        \begin{equation}
            \overline{I}_{\nu}^J = \int{\rm d}z\,\dfrac{c}{4\pi}\dfrac{\overline{j}_J(z)}{H(z)\,\Delta\nu\,(1+z)^2}\,\mathcal{R}_\nu\!\left(\dfrac{\nu_J}{1+z}\right)\equiv\int{\rm d}z\,\xi^J_\nu(z) ,
        \end{equation}
        which defines the kernel of Eq.~\eqref{eq:emiss_kernel}. For the top-hat bandpasses used in this work, $\mathcal{R}_\nu=1$ for $\nu_{\rm min}\leq\nu_{\rm obs}\leq\nu_{\rm max}$ and zero otherwise, so the kernel is non-zero only in the redshift range of Eq.~\eqref{eq:zrange}.

        With $\overline{j}_J$ in $L_\odot\,{\rm Mpc}^{-3}$, $c/H$ in Mpc, and $\Delta\nu$ in Hz, the kernel is in $L_\odot\,{\rm Mpc}^{-2}\,{\rm Hz}^{-1}\,{\rm sr}^{-1}$ per unit redshift. It is converted to ${\rm Jy\,sr^{-1}}$ with $1\,L_\odot\,{\rm Mpc}^{-2}\,{\rm Hz}^{-1}=4.02\times10^{7}$\,Jy. Finally, the spectra are converted from ${\rm Jy^2\,sr^{-1}}$ to $\mu{\rm K}^2_{\rm CMB}$ by dividing by $b_\nu b_{\nu^\prime}$, where
        \begin{equation}
            b_\nu=\dfrac{1}{\Delta\nu}\int{\rm d}\nu_{\rm obs}\,\mathcal{R}_\nu(\nu_{\rm obs})\left.\dfrac{\partial B_{\nu_{\rm obs}}(T)}{\partial T}\right|_{T_{\rm CMB}}
        \end{equation}
        is the band-averaged derivative of the Planck function at the CMB temperature.

    \subsection{CIB}
        For a continuum source with a comoving specific luminosity density $\epsilon_{\nu_{\rm rest}}(z)$, the observed specific intensity is
        \begin{equation}
            I_\nu(\nu_{\rm obs})=\int{\rm d}z\,\dfrac{c}{4\pi H(z)}\,\dfrac{\epsilon_{(1+z)\nu_{\rm obs}}(z)}{1+z} .
        \end{equation}
        Averaging over the channel and changing variable to the rest-frame frequency, $\nu_{\rm rest}=(1+z)\nu_{\rm obs}$, gives
        \begin{equation}
            \overline{I}_\nu=\int{\rm d}z\,\dfrac{c}{4\pi}\dfrac{1}{H(z)\,(1+z)^2\,\Delta\nu}\int{\rm d}\nu_{\rm rest}\,\mathcal{R}_\nu\!\left(\dfrac{\nu_{\rm rest}}{1+z}\right)\epsilon_{\nu_{\rm rest}}(z) ,
        \end{equation}
        which is Eq.~\eqref{eq:cib_kernel} with
        \begin{equation}
            \overline{j}_{{\rm CIB},\nu}(z)=\int{\rm d}\nu_{\rm rest}\,\mathcal{R}_\nu\!\left(\dfrac{\nu_{\rm rest}}{1+z}\right)\epsilon_{\nu_{\rm rest}}(z) ,
        \end{equation}
        i.e. the infrared luminosity density emitted in the rest-frame band $[(1+z)\nu_{\rm min},(1+z)\nu_{\rm max}]$. With this definition, the CO and CIB kernels have the same form, and, for a line ($\epsilon_{\nu_{\rm rest}}=\overline{j}_J\,\delta_{\rm D}(\nu_{\rm rest}-\nu_J)$), the CIB expression reduces to the CO one. For the CIB, each galaxy emits a specific luminosity $L_{\rm IR}\Theta(\nu_{\rm rest})$, with $L_{\rm IR}=\kappa_{\rm IR}\rm{SFR}$ and $\Theta\propto\nu^\beta B_\nu(T_d)$ a modified black body normalised to unity, with $T_d=34$K and $\beta=1.6$.

\section{Contribution of satellite galaxies}\label{app:co_gal}
    Previous works, mainly in the context of LIM experiments, have modelled the CO power spectrum using central galaxies only~\citep{Bernal:2019jdo, MoradinezhadDizgah:2021dei}. We therefore evaluate the impact of neglecting satellite galaxies on both the CO auto-spectrum and the \cocib\,cross-spectrum.

    We compute the ratio between the spectra obtained with central and satellite galaxies and those obtained with central galaxies only, for all \lact\,auto- and cross-frequency channels. Satellites increase the \coco\,power by $42-52\%$ for $\ell\gtrsim2000$ in all channels. The effect on \cocib\,is smaller: $24-33\%$ at $\ell\gtrsim2000$. Both effects are well below the modelling uncertainties discussed in Sect.~\ref{sec:mod_ass}. Neglecting satellites also roughly halves the computational time. The code therefore allows the user to compute the spectra either with the full model or with central galaxies only; the latter underestimates the \coco\,and the \cocib\,power by $\simeq30\%$ and $\simeq20\%$, respectively. In the Fisher forecasts of Sect.~\ref{sec:results}, we use the full model as our baseline and also consider the central-only case (case B).

\section{Foreground model}\label{app:fg_mod}
    In this appendix, we describe the model for the extragalactic foregrounds other than CO that enter the Fisher forecasts of Sect.~\ref{sec:results}. All the spectra are expressed in \muk\,and computed for the top-hat bandpasses of Table~\ref{tab:act_features}. The fiducial values of the model parameters are listed in Table~\ref{tab:fg_params}.

    \paragraph{CIB.} The CIB is described by the halo model of Sect.~\ref{subsec:cocib}, with the same SFR, CIB HOD, SED, and SFR scatter as used for the \cocib\,power spectrum: its 1h, 2h, and shot noise terms follow from Eqs~\eqref{eq:P1h_coco}--\eqref{eq:Pshot_coco}, with the CO luminosities replaced by the infrared luminosities and the CO HOD by the CIB HOD, and with the kernel of Eq.~\eqref{eq:cib_kernel}.

    \paragraph{tSZ.} The tSZ power spectrum is computed in the halo model as
    \begin{equation}
        C_{\ell,\nu\times\nu^\prime}^{\rm tSZ} = g(x_\nu)g(x_{\nu^\prime})T_{\rm CMB}^2\int \d z\dfrac{\d ^2V}{\d z\d\Omega}[Y_\ell^{1h}(z)+Y_\ell^{2h}(z)],
    \end{equation}
    with
    \begin{align}
        Y_\ell^{1h}(z) &= \int \d M\dfrac{\d n}{\d M}|y_\ell(M,z)|^2 ,\\
        Y_\ell^{2h}(z) &= P_{mm}^{\rm lin}(k,z)\biggl[\int \d M\dfrac{\d n}{\d M}b(M)y_\ell(M,z)\biggr]^2 ,
    \end{align}
    where $g(x)=x \rm{coth}(x/2)-4$ is the tSZ spectral function, $x_\nu=h\nu/(k_BT_{\rm CMB})$, and $y_\ell$ is the two-dimensional Fourier transform of the Compton-y profile of a halo of mass $M=M_{500}$.

    \paragraph{tSZ$\times$CIB.} The correlation between the tSZ and the CIB is computed with the 1h and 2h terms obtained by replacing one of the two $y_\ell$ factors of the tSZ expression with the CIB galaxy profile and its kernel.

    \paragraph{kSZ.} The kSZ power is described by the template $\mathcal{T}_\ell$ of~\cite{George:2014oba}, which includes the homogeneous and patchy reionisation contributions, rescaled to the amplitude $\mathcal{A}_{\rm kSZ}$ at $\ell=3000$: $D_\ell^{\rm kSZ}=\mathcal{A}_{\rm kSZ}\mathcal{T}_\ell/\mathcal{T}_{3000}$. In CMB units, it is the same in all channels. Replacing it with the late-time template of~\cite{Battaglia:2010tm}, adopted as a baseline by~\cite{Tristram:2025}, changes the parameter biases by less than $0.2\sigma$ and the uncertainties by less than 10\%. The conclusions of this work are stable against these two different choices of kSZ template.

    \paragraph{Radio sources.} Radio point sources are described by a Poisson term,
    \begin{equation}
        D_{\ell,\nu\times\nu^\prime}^{\rm radio} = \mathcal{A}_{\rm radio}\biggl(\dfrac{\ell}{3000}\biggr)^2\biggl(\dfrac{\nu\nu^\prime}{\nu_0^2}\biggr)^{\alpha_{\rm radio}}\epsilon_\nu\epsilon_{\nu^\prime} ,
    \end{equation}
    with $\nu_0=150.2$\ghz\, and $\epsilon_\nu$ the conversion factor from flux density to CMB temperature units.

    \paragraph{CMB.} The primary CMB power spectrum is computed with \camb\,for the cosmological parameters reported in Table~\ref{tab:fg_params}.

    \begin{table}
    \centering
    \caption{Fiducial values of the parameters varied in the Fisher forecasts.}
    \label{tab:fg_params}
    \begin{tabular}{lll}
        \hline\hline
        Parameter & Fiducial & Description \\
        \hline
        $\mathcal{A}_s$ & $2.101\times10^{-9}$ & Scalar amplitude \\
        $\Omega_m$ & 0.3097 & Matter density parameter \\
        $\Omega_b$ & 0.04897 & Baryon density parameter \\
        $\tau$ & 0.0544 & Reionisation optical depth \\
        $n_s$ & 0.9649 & Scalar spectral index \\
        $h$ & 0.6766 & Reduced Hubble constant \\
        \hline
        $\mathcal{A}_{\rm radio}$ & 1.11 & Radio amplitude [$\mu$K$^2$] \\
        $\alpha_{\rm radio}$ & -0.945 & Radio spectral index \\
        $\mathcal{A}_{\rm kSZ}$ & 1.9 & kSZ amplitude [$\mu$K$^2$] \\
        $b_{HSE}$ & 0.53 & Hydrostatic mass bias \\
        \hline
        $\log M_{\rm min}^{\rm CO}$ & 11.5 & Minimum mass of CO centrals \\
        $\sigma_{\rm CO}$ & 0.4 & Width of CO central occupation \\
        $M_{\rm sat}^{\rm CO}$ & 33.0 & Satellite mass scale (CO)\\
        $\alpha_{\rm CO}$ & 0.8 & Satellite slope (CO) \\
        \hline
        $\log M_{\rm min}^{\rm CIB}$ & 11.0 & Minimum mass of CIB centrals \\
        $\sigma_{\rm CIB}$ & 0.1 & Width of CIB central occupation \\
        $M_{\rm sat}^{\rm CIB}$ & 80.0 & Satellite mass scale (CIB)\\
        $\alpha_{\rm CIB}$ & 1.0 & Satellite slope (CIB) \\
        \hline
    \end{tabular}
\end{table}

\section{Fisher forecasts: full results}\label{app:fisher}
\begin{figure}
        \centering
        \includegraphics[width=\linewidth]{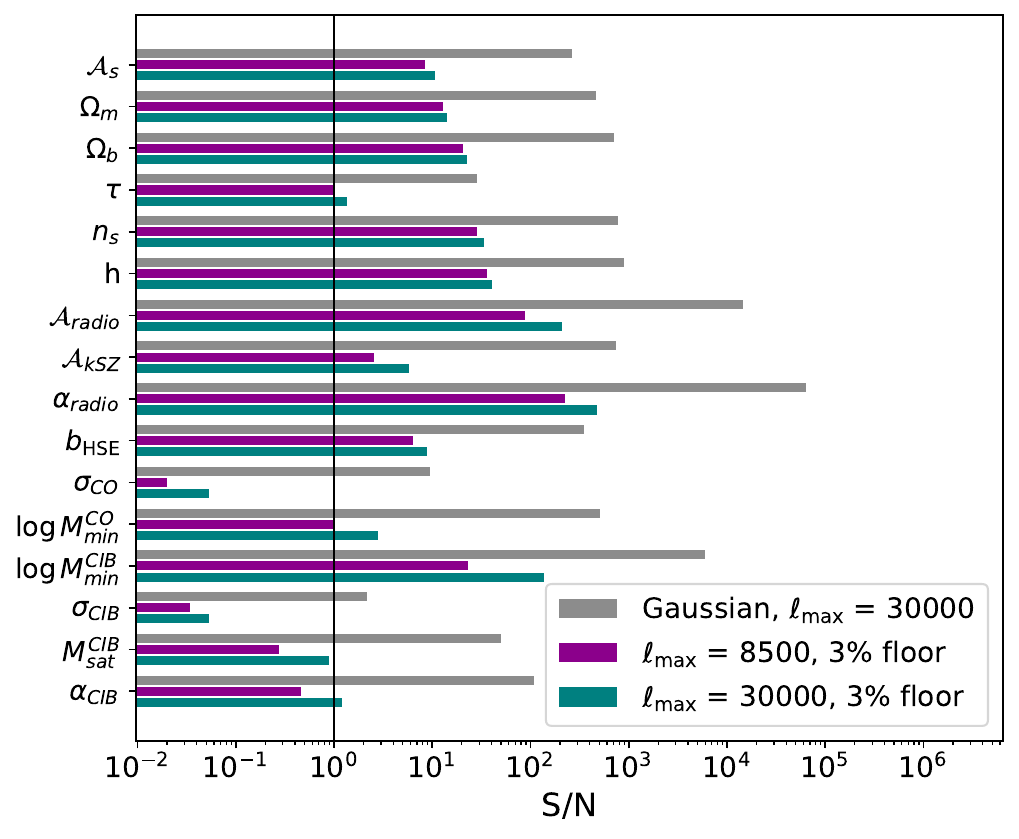}
        \caption{Same as the right panel of Fig.~\ref{fig:fisher_hd_A}, but for case B (CO from central galaxies only).}
        \label{fig:snr_hd_nosat}
    \end{figure}

\begin{figure*}
        \centering
        \includegraphics[width=0.49\linewidth]{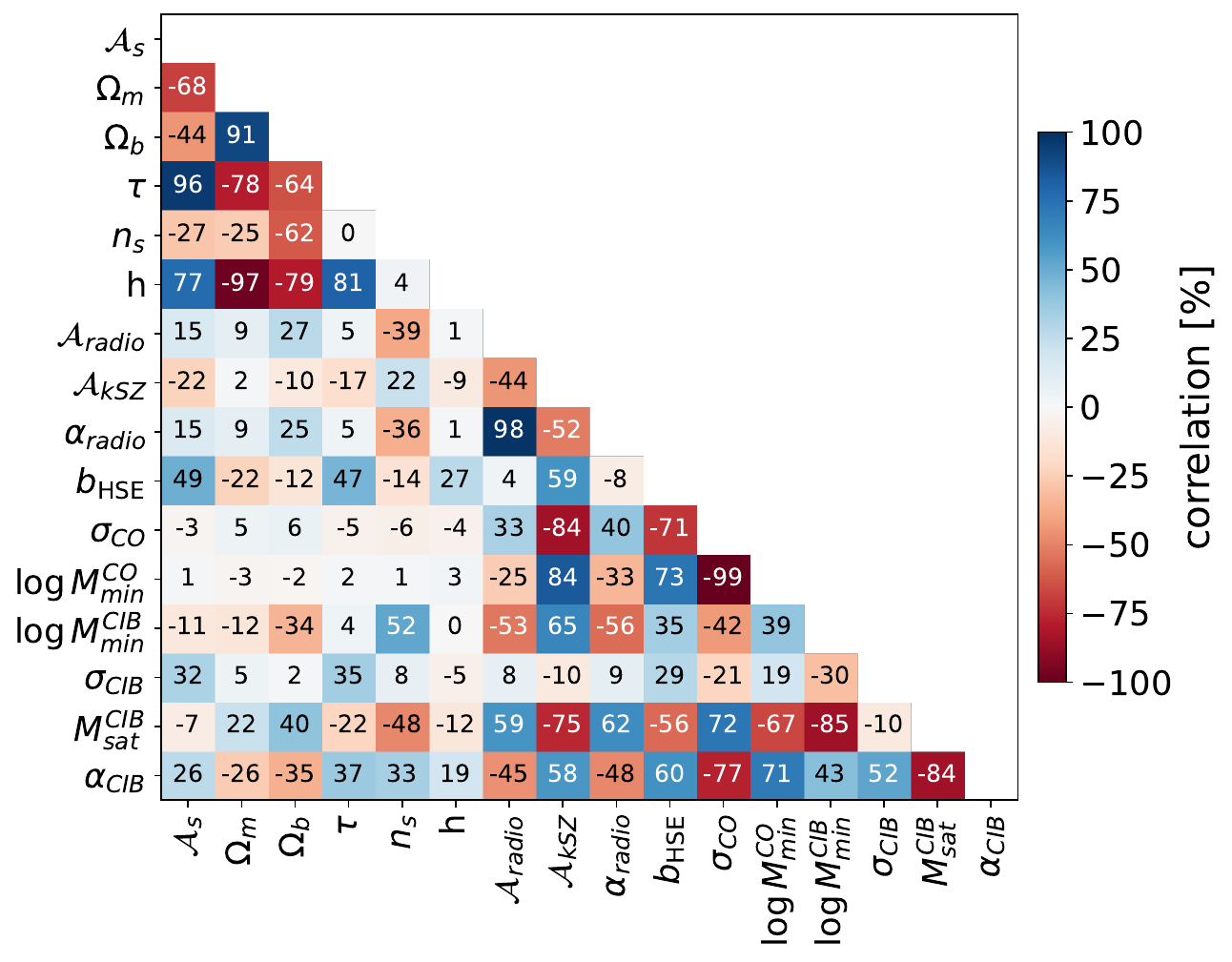}
        \includegraphics[width=0.49\linewidth]{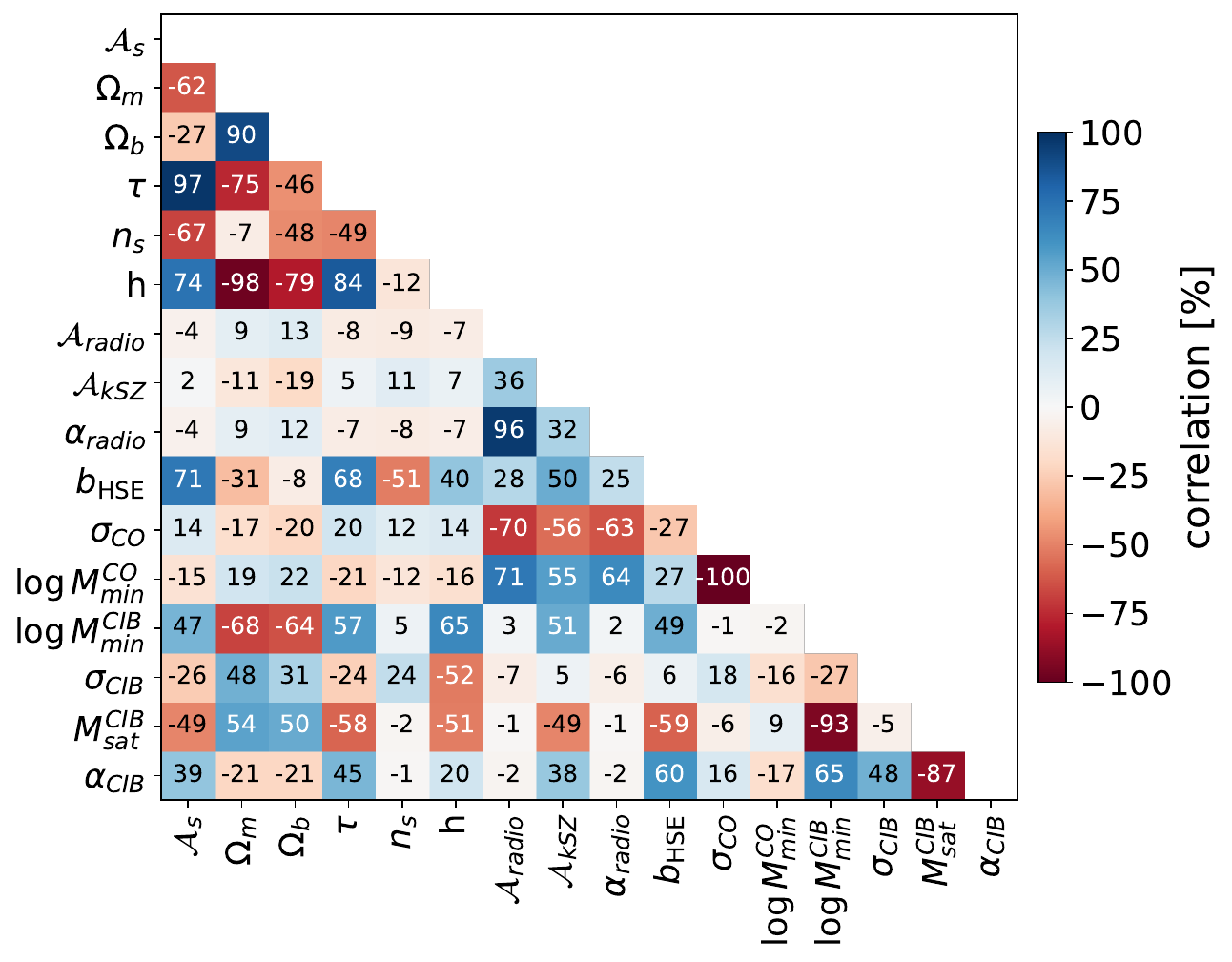}
        \caption{Correlation matrices for the \lact\,configuration (left) and for the \lcmnhd\,configuration with the 3\% floor and $\ell_{\rm max}=30000$ (right) when no CO satellite galaxies are included in the model (case B).}
        \label{fig:app_corr_nosat}
    \end{figure*}

    In this appendix, we show the results of the Fisher forecasts of Sect.~\ref{sec:results} that are not included in the main text. Figure~\ref{fig:app_corr_nosat} shows the correlation matrices for case B (no CO satellite galaxies in the model) for the \lact\,(left) and \lcmnhd\,(right) configurations; as in the main text, the \lcmnhd\,matrix is computed with the 3\% systematic error floor and $\ell_{\rm max}=30000$. Figure~\ref{fig:snr_hd_nosat} shows the SNR of case B for the \lcmnhd\,configuration, with the Gaussian covariance and with the 3\% floor for the two values of $\ell_{\rm max}$.

\begin{figure*}
        \centering
        \includegraphics[width=0.49\linewidth]{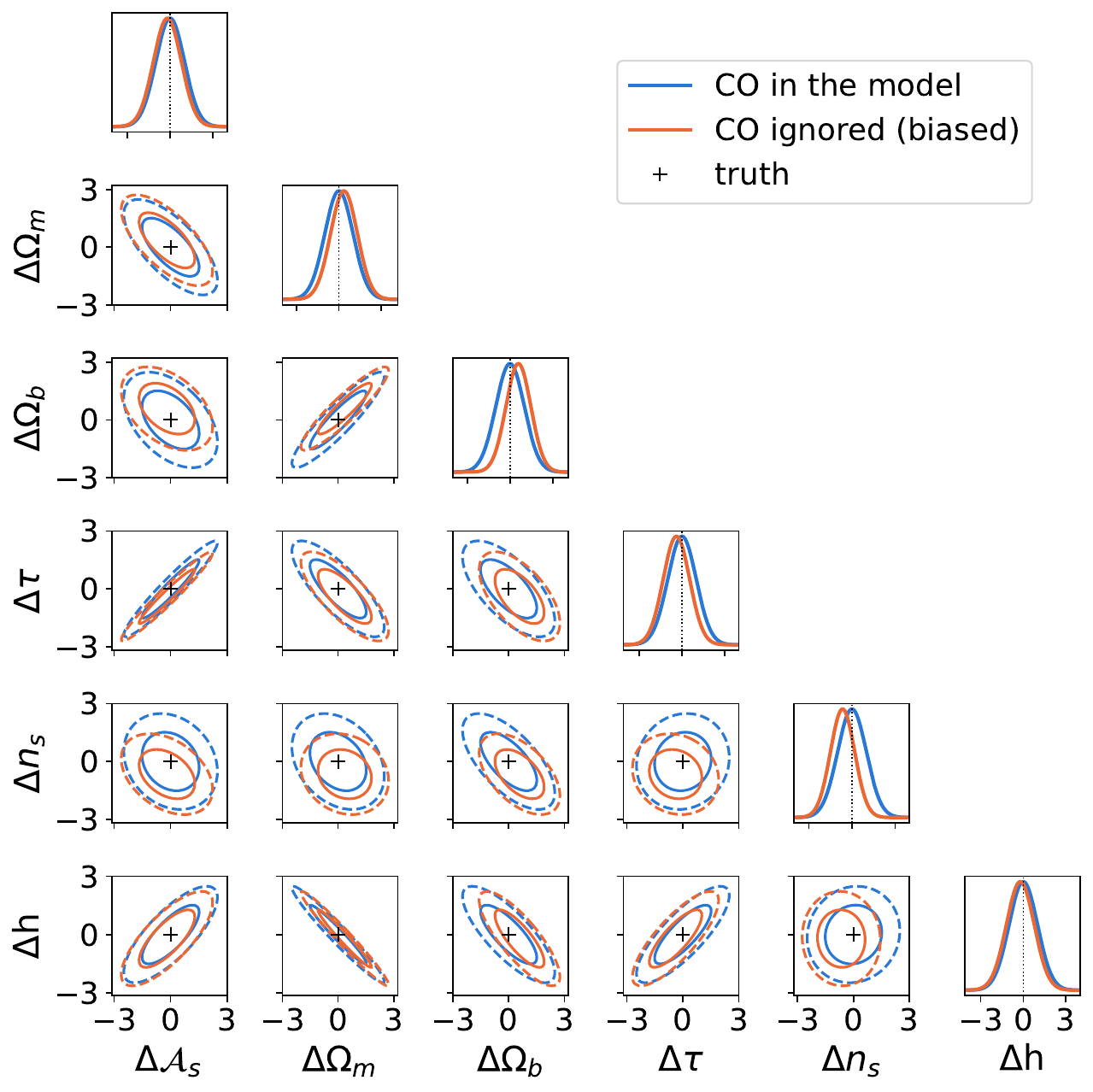}\hfill
        \includegraphics[width=0.49\linewidth]{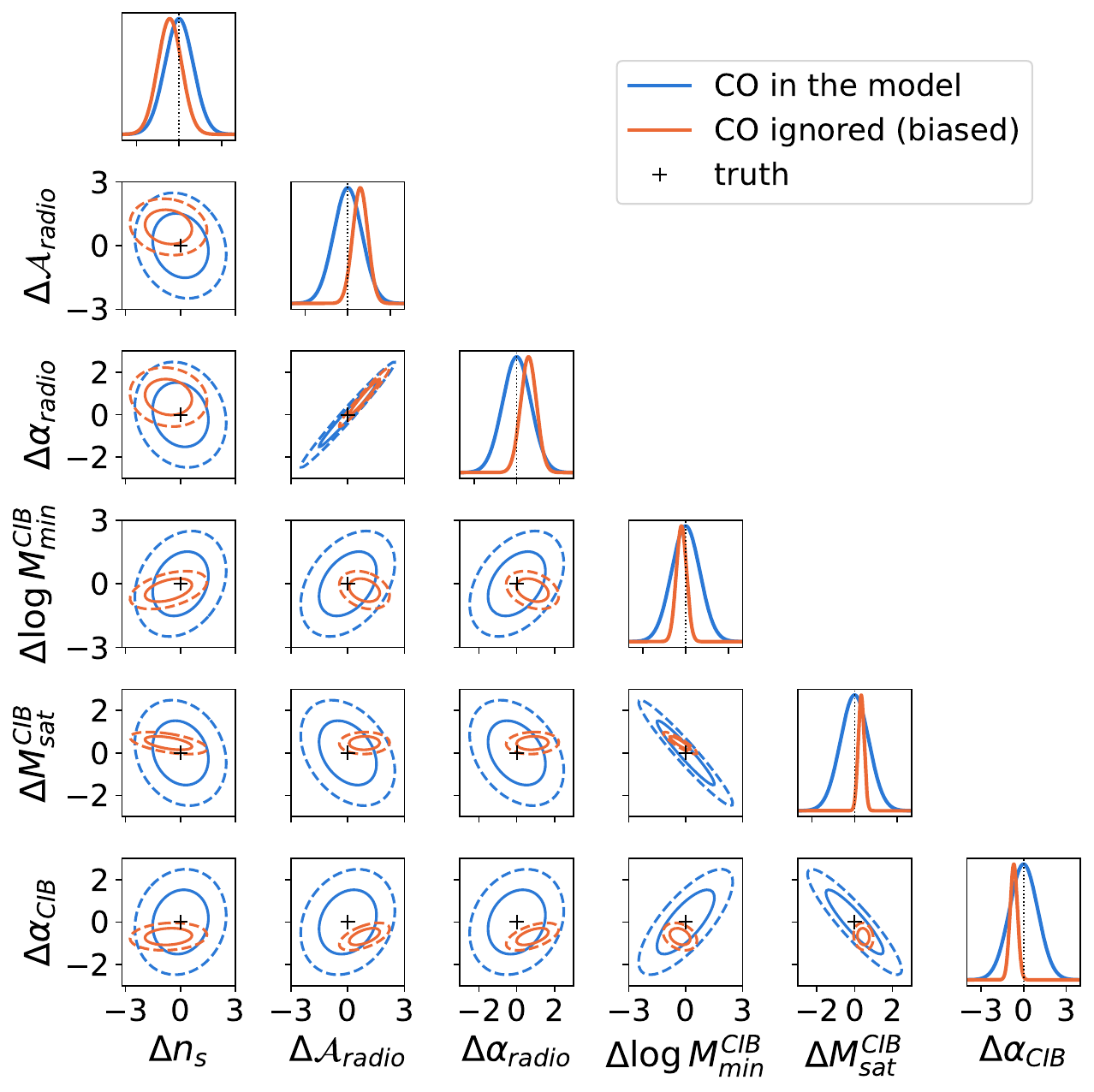}
        \caption{Forecast constraints for the \lact\,configuration (case A) with CO in the model (blue) and with CO neglected (orange, shifted by the bias of Fig.~\ref{fig:bias_act_hd}), for the cosmological parameters (\textit{left}) and the foreground parameters (\textit{right}).}
        \label{fig:act_triangles}
    \end{figure*}

    Figure~\ref{fig:act_triangles} compares the \lact\,constraints with and without CO in the model for the cosmological parameters (left) and the foreground parameters (right). Table~\ref{tab:fisher_errors} reports the marginalised uncertainties for all the cases.

\begin{figure*}
        \centering
        \includegraphics[width=0.49\linewidth]{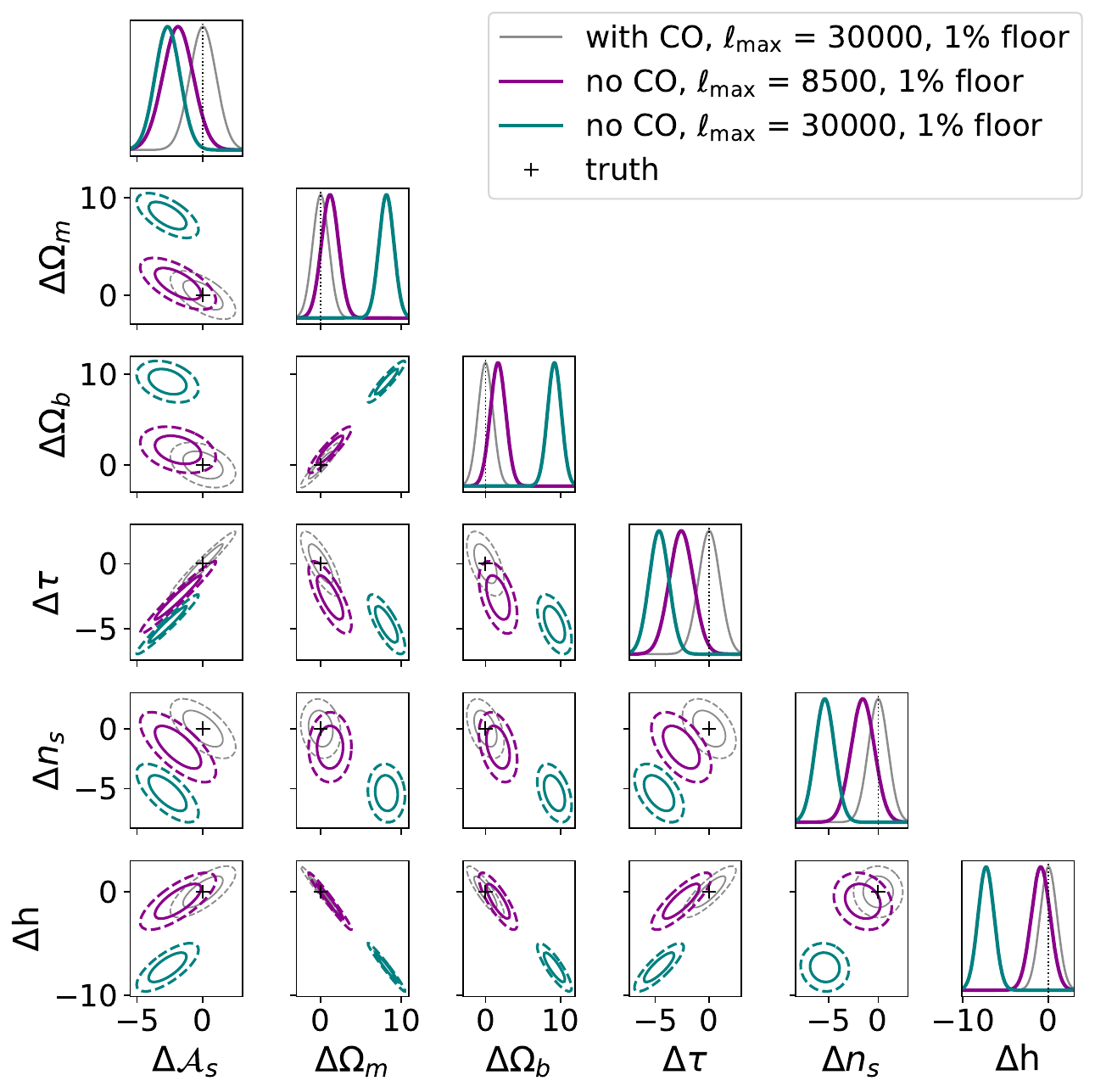}
        \includegraphics[width=0.49\linewidth]{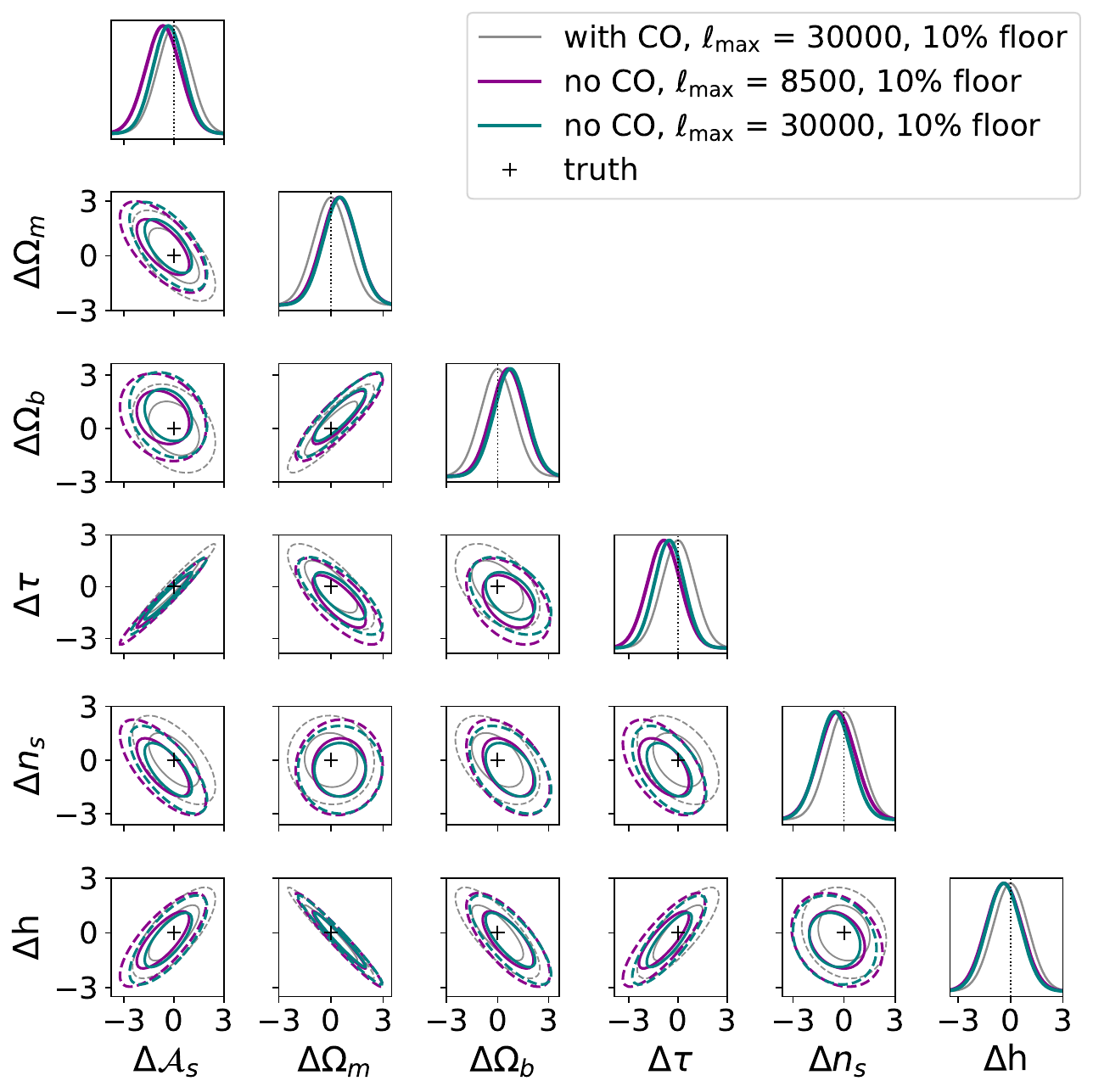}
        \caption{Same as Fig.~\ref{fig:hd_lmax_triangle}, but for a 1\% systematic error floor (\textit{left panel}) and for a 10\% systematic error floor (\textit{right panel}).}
        \label{fig:hd_bias_1pc}
    \end{figure*}

    Finally, Fig.~\ref{fig:hd_bias_1pc} shows the cosmological constraints obtained by neglecting CO in the \lcmnhd\,forecast with a 1\% systematic error floor (left) and with a 10\% error floor (right).

\begin{table*}
    \centering
    \caption{Marginalised $1\sigma$ uncertainties on the parameters for the ACT-like and CMB-HD-like configurations, with cosmological and foreground parameters varied.}
    \label{tab:fisher_errors}
    \begin{tabular}{l|ccc|ccc}
        \hline\hline
        & \multicolumn{3}{c|}{ACT-like} & \multicolumn{3}{c}{CMB-HD-like} \\
        Parameter & Baseline & Centrals only & no CO & Gaussian & 3\% floor & 3\% floor, no CO \\
        \hline
        $\mathcal{A}_s$ & $1.6\times10^{-10}$ & $1.5\times10^{-10}$ & $1.5\times10^{-10}$ & $9.0\times10^{-12}$ & $2.0\times10^{-10}$ & $1.9\times10^{-10}$\\
        $\Omega_m$ & 0.017 & 0.017 & 0.016 & $6.5\times10^{-4}$ & 0.021 & 0.021\\
        $\Omega_b$ & 0.0019 & 0.0018 & 0.0017 & $7.0\times10^{-5}$ & 0.0021 & 0.0020\\
        $\tau$ & 0.037 & 0.035 & 0.034 & 0.0020 & 0.042 & 0.039\\
        $n_s$ & 0.025 & 0.024 & 0.021 & 0.0015 & 0.028 & 0.028\\
        $h$ & 0.013 & 0.013 & 0.013 & $7.8\times10^{-4}$ & 0.017 & 0.016\\
        \hline
        $\mathcal{A}_{\rm radio}$ & 0.096 & 0.075 & 0.052 & $7.4\times10^{-5}$ & 0.0045 & 0.0035\\
        $\alpha_{\rm radio}$ & 0.068 & 0.054 & 0.038 & $1.5\times10^{-5}$ & 0.0017 & 0.0015\\
        $\mathcal{A}_{\rm kSZ}$ & 0.69 & 0.69 & 0.37 & 0.0026 & 0.40 & 0.27\\
        $b_{HM}$ & 0.063 & 0.063 & 0.043 & 0.0017 & 0.060 & 0.056\\
        \hline
        $\log M_{\rm min}^{\rm CO}$ & 9.0 & 11 & -- & 0.0084 & 2.7 & --\\
        $\sigma_{\rm CO}$ & 33 & 24 & -- & 0.019 & 4.6 & --\\
        $M_{\rm sat}^{\rm CO}$ & $8.9\times10^3$ & -- & -- & 4.4 & $4.1\times10^2$ & --\\
        $\alpha_{\rm CO}$ & $1.7\times10^2$ & -- & -- & 0.095 & 8.6 & --\\
        \hline
        $\log M_{\rm min}^{\rm CIB}$ & 1.3 & 0.56 & 0.46 & 0.0018 & 0.13 & 0.066\\
        $\sigma_{\rm CIB}$ & 2.2 & 2.1 & 2.0 & 0.051 & 1.9 & 1.8\\
        $M_{\rm sat}^{\rm CIB}$ & $1.0\times10^3$ & $4.6\times10^2$ & $2.1\times10^2$ & 1.6 & $1.4\times10^2$ & 75\\
        $\alpha_{\rm CIB}$ & 5.6 & 3.7 & 1.5 & 0.011 & 1.2 & 0.75\\
        \hline
    \end{tabular}
    \tablefoot{``Baseline'': CO emitted by central and satellite galaxies (case A). ``Centrals only'': CO emitted by central galaxies only (case B). ``no CO'': CO included in the covariance but not in the model (Sect.~\ref{subsec:noco}). For the CMB-HD-like configuration, all the columns refer to case A and $\ell_{\rm max}=30000$; ``Gaussian'': covariance of Eq.~\eqref{eq:covmat}; ``3\% floor'': with the systematic error floor of Sect.~\ref{subsub:cmbhdL}.}
\end{table*}

\end{document}